\documentclass[]{jfm}
\usepackage{graphicx}
\usepackage{epstopdf, epsfig}
\usepackage{hyperref}
\usepackage{verbatim}

\usepackage{amsmath,amssymb,bm}
\usepackage{caption}
\usepackage{subcaption}
\usepackage{breqn}
\usepackage{xcolor}
\usepackage{cleveref}
\usepackage{ulem}

\shorttitle{Multicomponent vesicles under shear flow}
\shortauthor{A. Venkatesh and V. Narsimhan}

\title{Shape dynamics of nearly spherical, multicomponent vesicles under shear flow}

\author{Anirudh Venkatesh\aff{1}
 \and Vivek Narsimhan \aff{1} \corresp{\email{vnarsim@purdue.edu}}
 }

\affiliation{\aff{1}Davidson School of Chemical Engineering, Purdue University, IN 47906, USA
}

\begin{document}

\maketitle

\begin{abstract}
In biology, cells undergo deformations under the action of flow caused by the fluid surrounding them. These flows lead to shape changes and instabilities that have been explored in detail for single component vesicles. However, cell membranes are often multi-component in nature, made up of multiple phospholipids and cholesterol mixtures that give rise to interesting thermodynamics and fluid mechanics. Our work analyses linear flows around a multi-component vesicle using a small-deformation theory based on vector and scalar spherical harmonics. We set up the problem by laying out the governing momentum equations and the traction {balance } arising from the phase separation and bending. These equations are solved along with a Cahn-Hilliard equation that governs the coarsening dynamics of the phospholipid-cholesterol mixture. We provide a detailed analysis of the vesicle dynamics (e.g., tumbling, breathing, tank-treading, swinging, and phase treading) in two regimes -- when flow is faster than coarsening dynamics (Peclet number $Pe \gg 1$) and when the two time scales are comparable ($Pe \sim O(1)$) -- and provide a discussion on {when these behaviours occur}. The analysis aims to provide an experimentalist with important insights pertaining to the phase separation dynamics and their effect on the deformation dynamics of a vesicle.

\end{abstract}

\begin{keywords}
Authors should not enter keywords on the manuscript, as these must be chosen by the author during the online submission process and will then be added during the typesetting process (see http://journals.cambridge.org/data/\linebreak[3]relatedlink/jfm-\linebreak[3]keywords.pdf for the full list)
\end{keywords}

\section{Introduction}\label{sec:intro}

Flow around biological membranes holds a great deal of importance in a multitude of systems \citep{seifert1999fluid,Herzenberg2006,Shi2018,Salmond2021}. These biological membranes are modeled using surrogate structures known as vesicles. Vesicles are complex droplets having a lipid bilayer on the surface instead of simple fluid boundaries. The presence of this lipid bilayer imparts an elastic resistance to bending \citep{Helfrich1973}. These lipid bilayer membranes act like 2-dimensional fluid sheets that resist any stretching or compression \citep{campelo2014helfrich}. Apart from being a marvelous surrogate system for {understanding the biophysics of cellular membranes}, lipid vesicles are often used as carriers for drug delivery processes \citep{guo2003chemical,Needham_1999}. 

Lipid bilayers are made up for long chain compounds known as phospholipids that contain a hydrophillic head and a hydrophobic tail \citep{alberts2017molecular}. When the bilayer consists of a single type of phospholipid, we call them `single component' vesicles. These vesicles have been the centre of a plethora of studies over the past 5 decades. The physical properties and thermodynamics of such vesicles can be characterized by a lipid bilayer exhibiting a uniform bending resistance \citep{lipowsky1995structure}. Previous studies have explored the dynamics of such single component vesicles in great detail by studying multiple modes of vesicle motion in different systems (a) shear flow -- tank-treading \citep{keller1982motion,seifert1999fluid,abkarian2002tank}, swinging \citep{Noguchi2007}, tumbling \citep{biben2003tumbling,rioual2004analytical,noguchi2005dynamics,kantsler2006transition}, and vacillated breathing \citep{misbah2006vacillating} (b) extensional flow \citep{zhao2013shape,narsimhan2014mechanism,narsimhan2015pearling,dahl2016experimental,kumar2020conformational} (c) general linear flows \citep{Vlahovska2007,Lin2019General} (d) oscillatory flows \citep{lin2021vesicle}. 

The deep understanding of single component vesicles has provided a base for further explorations into systems that are closer to reality. Often, these lipid bilayer membranes contain multiple phospholipids along with cholesterol molecules interspersed between them \citep{john2002transbilayer}. The existence of multiple molecules in these bilayers makes for an incredible amalgamation of phase equilibrium thermodynamics and fluid mechanics \citep{safran2018statistical}. Some of these phospholipids (e.g., dipalmitoylphosphatidylcholine(DPPC)) have a larger affinity towards cholesterol molecules than others (e.g., 1,2-Dioleoyl-sn-glycero-3-phosphocholine (DOPC)) \citep{veatch2005miscibility,davis2009phase,uppamoochikkal2010orientation}. This preferential separation into phases leads to the formation of ordered and disordered liquid phases on the membrane surface \citep{veatch2005seeing}, a term known as `lipid rafts' \citep{simons1997functional}. These rafts have a relevance in signal transduction \citep{simons2000lipid} and protein transfer across membranes, thereby having implications for health and diseases \citep{michel2007lipid}. These factors underscore the importance of studying such systems. 

From a mechanical viewpoint, the existence of multiple phases imparts inhomogeneous properties like the bending rigidity to  the vesicle due to the differences in bending stiffnesses of each constituent phospholipid \citep{baumgart2003imaging}. Moreover, the resultant lateral phases fall prey to a tussle between convective motion due to the background fluid and surface diffusive motion due to the inherent molecular properties of the phases \citep{yanagisawa2007growth,taniguchi2011numerical,arnold2023active}. 

While creating medical diagnostic devices, often, the measurement of mechanical properties of such vesicles is important \citep{kollmannsberger2011,lei2021cancer}. These measurements help in improving the control and precision of devices. Previously, multiple numerical simulations have been performed to understand the motion of multicomponent vesicles under shear flow \citep{sohn2010dynamics,liu2017dynamics,gera2018three}. These studies have highlighted the influence of line tension and bending rigidity, along with the membrane tension of the vesicle, on the modes of motion that the vesicle undergoes -- tank treading, tumbling, phase treading, vertical banding, among others. More recently, under the limit of dominant advective forces, authors came up with an analytical treatment of two-dimensional multicomponent vesicles and their swinging to tumbling transition \citep{gera_salac_spagnolie_2022}. The transition primarily depended on a ratio of the bending stiffness of the two phases and the capillary number of the vesicle. While informative and thoughtful, the study lacked an analytical treatment of the case when diffusive timescales are comparable to that of the convective timescales. Secondly, the study treated a multicomponent vesicle as a two-dimensional inexstensible membrane, thus leaving room for out-of-plane deformations. To the authors' knowledge, such an analytical treatment is has not been provided yet. 

We aim to bridge this gap of knowledge through this study that focuses on the semi-analytical prediction for a dilute, three-dimensional, nearly-spherical, multicomponent vesicle suspension placed under a background shear flow. We leverage the vector spherical harmonics-based techniques previously used for single component vesicles and drops \citep{vlahovska2005deformation,Vlahovska2007} to solve the underlying non-linear dynamical equations up to leading order, while ensuring conservation of the composition and vesicle surface area. This helps us arrive at reduced order equations governing the shape and composition of the vesicle and the phospholipid-cholesterol phases respectively. We delineate multiple motions exhibited by the vesicle and discuss their dependence on material properties. The aim of this study is to afford an experimental researcher with a theory that could \textit{a priori} predict the vesicle dynamics based on the material specifications and control variables.

We discuss the mathematical formulation and problem setup in section \ref{sec:problem_formulation}. We then go over the approximations for a vesicle undergoing small deformations in the weakly-segregated limit in section \ref{sec:sph_harm_exp}. We refresh the memory of the reader by introducing them to previously studied single-component vesicles in section \ref{sec:single_component_ves}. We discuss the multicomponent vesicle results in section \ref{sec:multicomponent_results}. We then conclude the study with a discussion in section \ref{sec:discussion}.

\section{Problem formulation}\label{sec:problem_formulation}

\begin{figure}
    \centering
    \includegraphics[width=\linewidth]{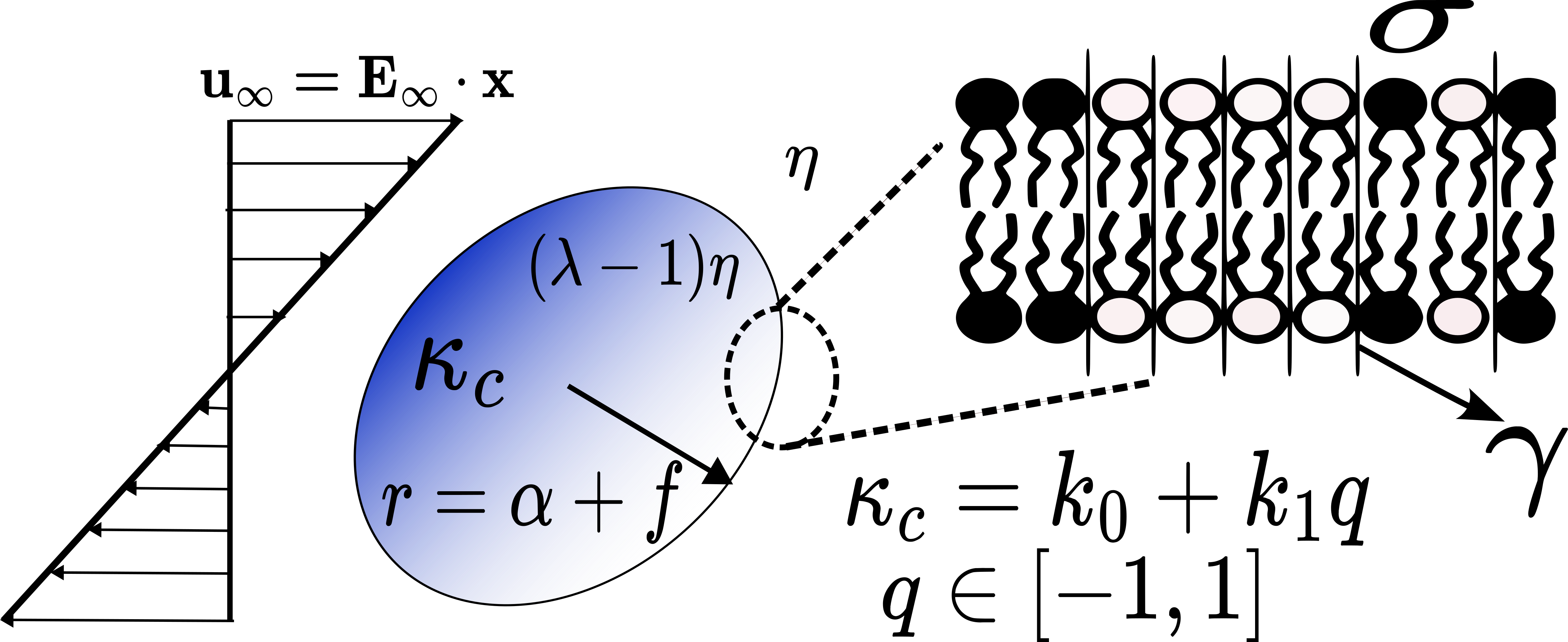}
    \caption{Schematic for a Newtonian fluid enclosed by a spherical multicomponent lipid bilayer separating the fluid from a surrounding Newtonian fluid. The inset figure shows a zoomed-in version of the lipid bilayer and its properties.}
    \label{fig:Schematic_Flow}
\end{figure}

We consider a lipid membrane vesicle containing a ternary mixture of phospholipids and cholesterol undergoing phase separation on the membrane surface (see figure \ref{fig:Schematic_Flow}). This vesicle is placed in an unbounded domain containing a background shear flow $\boldsymbol{u}^{\infty} = \dot{\gamma} y \boldsymbol{\hat{x}}$. This membrane contains a Newtonian fluid with viscosity $(\lambda-1) \eta$ and a surrounding fluid with viscosity $\eta$ separated by the bilayer surface. The bilayer has a bending stiffness of $\kappa_{c}$ that is dependent on the phospholipid distribution {characterized by an order parameter $q$}. This dependence will be explained in section \ref{sec:energy_mem_setup}. Furthermore, the lipid molecules impose a membrane tension to preserve the membrane surface area $A$ and there exists a line tension between the separated phases (indicated by black and white colored lipid heads). The vesicle has a characteristic size $R$ which is the radius of a sphere of same volume. 

\subsection{Free energy of membrane}\label{sec:energy_mem_setup}
The lipid membrane energy landscape has been a problem of interest for the past few decades. Historically, a lipid bilayer membrane is treated as an elastic surface. Multiple models have been used to describe such materials -- Keller Skalak model \citep{keller1982motion}, area difference model \citep{seifert1997configurations}, {and} Helfrich model \citep{Helfrich1973,campelo2014helfrich} to name a few. In our study, we use a Canham-Helfrich model.






The {free} energy {of a multicomponent membrane is governed by three factors: bending, mixing, and surface tension energies.} The bending energy is given by 

\begin{equation}\label{eq:bending_Helfrich}
    W_{bend} = \int{\frac{1}{2}\kappa_{c} (2H-c_{0})^{2}dS}
\end{equation}
{where $\kappa_{c} = k_{0} + k_{1}q$ is a bending modulus that depends on the lipid phase behaviour characterized by an order parameter $q$.  The order parameter $q$ is a variable whose values can range from $-1$ corresponding to pure $L_{d}$ (cholesterol-lacking) and $+1$ corresponding to pure $L_{o}$ (cholesterol-rich). Thermodynamically speaking, this is the local coordinate along a tie-line in the two-phase coexistence region that is present in the ternary phase diagram for a system like DOPC, DPPC, and chol.  The values of $k_0$ and $k_1$ are $k_0 = \frac{1}{2} \left( \kappa_{lo} + \kappa_{ld} \right)$ and $k_1 = \frac{1}{2} \left( \kappa_{lo} - \kappa_{ld} \right)$, where $\kappa_{l0}$ and $\kappa_{ld}$ are the bending moduli of the $L_0$ and $L_d$ phases, respectively. Lastly, the parameters $H$ and $c_{0}$ are the mean and spontaneous curvatures of the bilayer leaflet. We treat $c_{0}=0$ for a symmetric leaflet.}



The mixing energy is given by the {Landau-Ginzberg} equation \citep{gompper1992ginzburg}:
\begin{equation}\label{eq:mixing_energy}
W_{mix} = \int{\left(\frac{a}{2}q^{2}+\frac{b}{4}q^{4} + \frac{\gamma^{2}}{2}|\nabla_{s}q|^{2}\right)dS}
\end{equation}

The first two terms represent a double-well potential that gives rise to phase separation (for $a < 0$), and the last term is an energy penalty for creating phases that is related to line tension.  This free energy  has been used in many studies of lipid systems undergoing phase separation. These parameters are related to the experimentally-measured phase split concentrations ($q^{\pm}$), line tension $(\xi^{line})$, and the interface width $(\varepsilon^{width})$ as follows:
\begin{equation}\label{eq:linetension_equation}
    q^{\pm} = \pm \sqrt{|a|/b} \qquad \xi^{line} = \frac{2\sqrt{2}}{3b}|a|^{3/2}\gamma \qquad  \varepsilon^{width} = \sqrt{\frac{2\gamma^{2}}{|a|}}
\end{equation}

The Landau-Ginzberg equation has been used to qualitatively model bilayer membranes \citep{gera2018three} -- see appendix of \cite{camley2014fluctuating} for the estimated dependence of $a,b$, and $\gamma$ for a specific experimental system ($R\sim O(nm)$). 

Lastly, the surface tension energy is given by:

\begin{equation}
    W_{\sigma} = \int{\sigma dS} = \text{constant}
\end{equation}
The above contribution is constant because the lipid area per molecule is conserved and thus the membrane is incompressible.  The surface tension $\sigma$ thus acts as a Lagrange multiplier to ensure area conservation.

\subsection{Fluid motion and membrane interface dynamics}\label{sec:dyn_equations}






We solve the Stokes equations inside and outside the vesicle that govern the velocity, stress, and pressure distributions.

\begin{subequations}
\begin{equation}
    (\lambda-1)\eta\nabla^{2}\boldsymbol{u}^{in} = \nabla p^{in}
\end{equation}
\begin{equation}
    \eta\nabla^{2}\boldsymbol{u}^{out} = \nabla p^{out}
\end{equation}
\end{subequations}
These equations are subject to the following boundary conditions:\\

\begin{itemize}
    \item \underline{Far-field}:  $\boldsymbol{u}^{out} \rightarrow \boldsymbol{u}^{\infty}$ as $|\boldsymbol{x}| \rightarrow \infty$
    \item \underline{Continuity of velocity}:  $\boldsymbol{u}^{in} = \boldsymbol{u}^{out}$ for $\boldsymbol{x} \in S$
    \item \underline{Membrane incomressibility}:  $\nabla_s \cdot \boldsymbol u = 0$ for $\boldsymbol{x} \in S$
    \item \underline{Traction balance}:  $\boldsymbol{f}^{hyd} =  \boldsymbol{f}^{mem}$ for $\boldsymbol{x} \in S$\\
\end{itemize}

In the traction balance, $\boldsymbol{f}^{hyd} = \boldsymbol{n} \cdot \left(\boldsymbol{T}^{out} - \boldsymbol{T}^{in} \right)$ is the hydrodynamic traction from viscous stresses, where $\boldsymbol{n}$ is the outward-pointing normal vector and  $\boldsymbol{T} = \mu \left( \nabla \boldsymbol{u} + \nabla \boldsymbol{u}^T \right) - p \boldsymbol{I}$ is the stress tensor for a fluid with viscosity $\mu$ ($\mu = \eta$ inside the vesicle, $\mu = (\lambda -1) \eta$ outside the vesicle).  The membrane traction $\boldsymbol{f}^{mem}$ is equal to the first variation of the membrane free energy with respect to position:  $\boldsymbol{f}^{mem} = \delta W/\delta \boldsymbol{x}$.  This can be broken into bending, mixing, and surface tension contributions $\boldsymbol{f}^{mem} = \boldsymbol{f}^{bend}+\boldsymbol{f}^{mix}+ \boldsymbol{f}^{\sigma}$, with expressions for each shown below:

\begin{subequations}
\begin{multline}\label{eq:f_bend_vector}
    \boldsymbol{f}^{bend} = \frac{\delta W_{bend}}{\delta \boldsymbol{x}} = -\boldsymbol{n}\nabla^{2}_{s}(\kappa_{c}(2H)) + \kappa_{c}(2H)\left[\nabla_{s}(2H) + (2K-4H^{2})\boldsymbol{n} \right]\\
    +2H\left(\frac{1}{2}\kappa_{c}(2H)^{2}\right)\boldsymbol{n} - \nabla_{s}\left(\frac{1}{2}\kappa_{c}(2H)^{2}\right)
\end{multline}    
\begin{multline}\label{eq:f_mix_vector}
   \boldsymbol{f}^{mix} = \frac{\delta W_{mix}}{\delta \boldsymbol{x}}= \gamma^{2} \nabla_{s}\cdot(\nabla_{s}q\nabla_{s}q) + 2H\left(\frac{1}{2}\gamma^{2}|\nabla_{s}q|^{2} + g(q)\right)\boldsymbol{n} \\
   - \nabla_{s}\left(\frac{1}{2}\gamma^{2}|\nabla_{s}q|^{2} + g(q)\right)
\end{multline}
\begin{equation}\label{eq:f_sigma_vector}
\boldsymbol{f}^{\sigma} = \frac{\delta W_{\sigma}}{\delta \boldsymbol{x}} = 2\sigma H\boldsymbol{n} - \nabla_{s}\sigma
\end{equation}
\end{subequations}
    
The reader is directed to the following publications for details on how these equations are derived \citep{Gera2017,Napoli_2010}.  In the above expressions, $\nabla_s = ( \boldsymbol{I} - \boldsymbol{nn} ) \cdot \nabla$ is the surface gradient, and $g = \frac{a}{2} q^2 + \frac{b}{4}q^4$ is the double well potential in the mixing free energy (Eq \ref{eq:mixing_energy}). The quantities $K = \text{det}(\boldsymbol{L}) = C_{1}C_{2}$ and $H = \frac{1}{2} \text{tr}(\boldsymbol{L}) = \frac{C_{1}+C_{2}}{2}$ are the Gaussian and mean curvatures of the interface respectively, where $\boldsymbol{L} = \nabla_{s}\boldsymbol{n}$ is the surface curvature tensor.  The surface tension $\sigma$ is a Lagrange multiplier to enforce membrane incompressibility ($\nabla_s \cdot \boldsymbol{u} = 0$ on the interface), and hence must be solved in addition to the velocity and pressure fields.

In addition the the flow field, one must also solve for order parameter $q$ on the membrane surface.  The flow and coarsening behaviour of the order parameter satisfy a Cahn-Hilliard equation \citep{Gera2017}:

\begin{subequations}\label{eq:CahnHilliard_Dynamic}
\begin{equation}
\frac{\partial q}{\partial t} + \nabla_{s}\cdot(\boldsymbol{u} q) = \frac{\nu}{\zeta_0} \nabla^{2}_{s} \zeta 
\end{equation}
\begin{equation}
    \zeta = \frac{\delta W}{\delta q} = \left( aq + bq^{3} - \gamma^{2}\nabla^{2}_{s}q + \frac{k_{1}}{2}(2H)^{2} \right)
\end{equation}
\end{subequations}
where in the above expression, $\nu$ is the mobility of the phospholipids (units $m^2/s$) and $\zeta = \delta W/\delta q$ is the surface chemical potential (units of energy per area).  The above expression states that lipids are either convected along the surface or move via gradients in chemical potential. The value reference chemical potential $\zeta_0$ is given in \citep{Gera2017}.

Lastly, we impose the kinematic boundary condition to avoid any slip on the membrane surface.  If we parameterize the vesicle radius as $r = r_s(\boldsymbol{x}, t)$, the kinematic boundary condition is: 

\begin{equation}\label{eq:kinematic_bc_vector}
\frac{D}{Dt} \left(r - r_s \right)= 0 \qquad \boldsymbol{x} \in S
\end{equation}
where $\frac{D}{Dt} = \frac{\partial}{\partial t} + \boldsymbol{u} \cdot \nabla$ is the substantial derivative.

\subsection{Time scales and dimensionless quantities}\label{sec:dim_quantities}

The vesicle dynamics occur over three timescales. The first one is a bending timescale $t_{bend} = R^{3}\eta/k_0$ that denotes the time taken by a vesicle to restore its equilibrium configuration under the action of bending forces. The second timescale is {that of the flow $t_{\dot{\gamma}} = \dot{\gamma}^{-1}$} and the third represents the timescale for coarsening $t_{q} = R^{2}/\nu$.  We pick the same characteristic scales for physical quantities as previously used in flow studies for single component vesicles \citep{Vlahovska2007}. All lengths are non-dimensionalized by the equivalent radius $R$, all times by the flow time scale $t_{\dot{\gamma}}=\dot{\gamma}^{-1}$, and all velocities by $U_{flow} = R/t_{\dot{\gamma}} = R\dot{\gamma}$. All pressures and viscous stresses are scaled by $\eta \dot{\gamma}$, whereas the {membrane tractions $\boldsymbol{f}^{mem}$} are scaled by $k_{0}/R^{2}$. 

Table \ref{tbl:Physical_parameter_range} lists the set of physical parameters for this problem and their typical experimental values, while Table \ref{tbl:Dimensionless_parameter_range} lists the dimensionless numbers for this problem.  {The first three dimensionless numbers are ones that are also found in the single component vesicle literature – the viscosity ratio parameter $\lambda = (\mu^{in} + \mu^{out})/\mu^{out}$ between the inner and outer fluids, the capillary number $\chi = \eta R^3/k_0$ relating the viscous to bending forces on the membrane, and the dimensionless excess area $\Delta = A/R^2 - 4\pi$ representing the floppiness of the vesicle.  The remaining dimensionless numbers occur only in multicomponent vesicles.  Of these, the most important ones are the dimensionless bending stiffness difference between the two phases $\beta = k_1/k_0 = (\kappa_{lo} - \kappa_{ld})/(\kappa_{lo}+\kappa_{ld})$, the Cahn number $Cn = \gamma/\left(R\sqrt{\zeta_0}\right)$ (i.e., ratio of line tension energy to the energy scale of phase separation), the surface Peclet number $Pe = R^{2}\dot{\gamma}/\nu$ (i.e., ratio of coarsening time from diffusion to flow time), and the line tension parameter $\alpha = k_0/\gamma^2$ (ratio between bending energy to line tension energy).  Note:  the Cahn number can be re-expressed as the ratio between the interface thickness and vesicle size $Cn = \epsilon^{width}/(\sqrt{2}R)$.}
Furthermore, the average concentration $q_{0}$ affects the position along the local energy landscape. It also affects stresses arising from the phase energy (Eq. \ref{eq:f_mix_vector}).

\begin{table}
\centering
  \begin{tabular}{p{0.13\textwidth}p{0.35\textwidth}p{0.29\textwidth}p{0.17\textwidth}}
    Variable & Name & Order of Magnitude & Reference\\
    \hline
    $R$ & Equivalent radius of spherical vesicle & $\sim O(1) \mu m$ & \cite{Deschamps2009}\\
    $k_{0}$ & Average bending stiffness between lipids $1$ and $2$ & $O(10^{-19}-10^{-18})J$ & \cite{Amazon2013}\\
    $k_{1}$ & Half of bending stiffness difference between lipids $1$ and $2$ & $O(10^{-19})J$ & \cite{Amazon2013}\\
    $\nu$ & Mobility of phospholipids & $O(10^{-11})m^{2}/s$ & \cite{Mobility_Negishi}\\
    $\gamma$ & Line tension parameter& $O(10^{-9}) J^{1/2}$& \cite{Luo_Maibaum_2020}\\
    $\dot{\gamma}$ & Shear rate & $O(0.1-1)s^{-1}$& \cite{Deschamps2009}\\
    \hline
\end{tabular}%
\vspace{-1em}
  \caption{Physical parameter ranges and orders of magnitude} 
\label{tbl:Physical_parameter_range}
\end{table}

\begin{table}
\centering
  \begin{tabular}{p{0.23\textwidth}p{0.43\textwidth}p{0.2\textwidth}}
    Variable & Name & Value\\ 
    \hline
    $\lambda=(\mu^{in} +\mu^{out})/\mu^{out} $ & Viscosity ratio parameter & $O(1-10)$ \\
    $\chi = \eta R^{3} \dot{\gamma}/k_{0}$ & Capillary number & $O(0.01-1)$\\
    $\Delta = A/R^2 - 4\pi$ & Excess area & $O(0.01 - 0.1)$\\
    $\tilde{a} = a/\zeta_{0}$ & Dimensionless double well potential term & -1 \\
    $\tilde{b} = b/\zeta_{0}$ &  Dimensionless double well potential term & $1$ \\
    $\beta = k_{1}/k_{0}$ & Ratio of bending stiffnesses & $O(0.1-1)$  \\
    $Cn = \gamma/(R\sqrt{\zeta_{0}})$ & Cahn number & $O(0.1-1)$ \\
    $\alpha = k_{0}/\gamma^{2}$ & Ratio of bending stiffness to line tension & $O(1)$ \\
    
    $Pe = R^{2}\dot{\gamma}/\nu$ & Peclet number (coarsening timescale/flow timescale) & $O(1-1000)$\\
    $q_0$ & Average order parameter & $[-1, 1]$\\
    \hline
\end{tabular}%
\vspace{-1em}
  \caption{Dimensionless parameter ranges and orders of magnitude } 
\label{tbl:Dimensionless_parameter_range}
\end{table}

\section{Solution for a nearly spherical vesicle}\label{sec:sph_harm_exp}

\subsection{Overview}
We solve for the vesicle shape and composition as a function of time in the limit small excess area  – i.e., $\Delta \ll 1$.  We use $\epsilon=\Delta^{1/2}$ as a perturbation variable and solve the dynamical equations in section \ref{sec:dyn_equations} to leading order while enforcing conservation of volume and area to $O(\epsilon^{2})$.  Details of the solution methodology are in the following subsections.
Section \ref{sec:pertub_expansion} discusses the perturbation expansion of the shape, surface tension, and concentration fields.  Sections \ref{sec:solve_Stokes} and \ref{sec:solve_CH} discuss the steps to solve these fields.  Section \ref{sec:numerics} describes the numerical method of solution.

\subsection{Perturbation expansion} \label{sec:pertub_expansion}

We expand the vesicle shape and composition in terms of spherical harmonics. The non-dimensional vesicle radius is written as $r = r_s(\theta, \phi)$, where 
\begin{equation} \label{eq:radius}
    r_{s}(\theta, \phi) = 1 + \sum_{l=1}^{\infty} \sum_{m=-l}^{+l} f_{lm}Y_{lm}(\theta, \phi)  - \frac{1}{4\pi}\sum_{lm}f_{lm}f_{lm}^{*}
\end{equation}

In the above equation, $Y_{lm}(\theta, \phi)$ are spherical harmonics (Appendix A), while $f_{lm}$ are coefficients of $O(\Delta^{1/2})$ that have to be solved.  The last term is present to satisfy the volume conservation constraint $V = 4 \pi/3$ to $O(\Delta)$ \citep{seifert1999fluid}.

The surface tension is expanded as:
\begin{equation}
    \sigma = \sigma_{0}+ \sum_{l=1}^{\infty} \sum_{m=-l}^{+l} \sigma_{lm}Y_{lm}(\theta, \phi)
\end{equation}
where $\sigma_{lm}$ are spherical harmonic coefficients of $O(\Delta^{1/2})$ that are determined by solving the Stokes equations (see next section).  The isotropic component $\sigma_0$ is solved by applying the constraint that the total area does not vary as a function of time. Satisfying this constraint to $O(\Delta)$ yields

\begin{equation} \label{eq:area_constraint_1}
\sum_{lm} (l + 2)(l-1) f_{lm}f^{*}_{lm} = \Delta
\end{equation}

Lastly, we expand the order parameter $q$ in terms of spherical harmonics 
\begin{equation} \label{eq:order_param}
    q = q_{0} + \sum_{l=1}^{\infty}\sum_{m=-l}^{+l}q_{lm}Y_{lm}(\theta, \phi) - \frac{1}{2\pi}\sum_{lm}q_{lm}f_{lm}^{*}
\end{equation}

In the above equation, $q_0$ is the average order parameter set by the composition of the membrane, and $q_{lm}$ are the coefficients for the spatially varying portion.  It assumed that $q_{lm} \sim O(\Delta^{1/2}) \ll 1$ -- i.e., the weak-segregation approximation. This approximation has been used in multiple studies pertaining to co-polymer systems \citep{leibler1980theory,fredrickson1987,seul1995domain}, and has been applied to multicomponent vesicles in the past \citep{taniguchi1994phase,kumar1999modulated}. The underlying assumption is that the phase segregation from a particular equilibrium critical point is weak, leading to small perturbations in the phase field.  We believe this is a good starting point towards building a semi-analytical theory for phase separation over lipid bilayer vesicles under flow.  Lastly, the final term in the above equation is there so that the order parameter is conserved to $O(\Delta)$ -- i.e., $\int q dA = q_0 A$.

In the following subsections, it is useful to represent the normal vector, mean curvature, surface Laplacian of mean curvature, and Gaussian curvature to $O(f)$.  These quantities are:

\begin{equation}
\begin{split}
    &\boldsymbol{n} \approx \hat{\boldsymbol{r}} - \sum_{lm} r\nabla f_{lm} Y_{lm} \qquad H \approx 1 + \frac{1}{2}\sum_{lm} (l+2)(l-1) f_{lm} Y_{lm} \\
    &  \nabla_s^2 H \approx -\frac{1}{2}\sum_{lm} (l+2)(l-1) l (l+1) f_{lm} Y_{lm} \qquad K \approx H^2
\end{split}
\end{equation}

\subsection{Solving Stokes equations} \label{sec:solve_Stokes}

We solve the Stokes equations around the vesicle to determine how the shape coefficients $f_{lm}$ evolve over time.  This section provides a high level overview of the steps, with algebraic details provided in Appendix \ref{appB}.

On a sphere $r = 1$, the membrane tractions $\boldsymbol{f}^{mem}$ give rise to an isotropic pressure, which does not contribute to the shape dynamics.  Thus, we Taylor expand $\boldsymbol{f}^{mem}$ to $O(\Delta^{1/2})$ on the surface of a sphere, and then solve Stokes equations on the sphere with those boundary conditions.  We note that this idea is commonplace in studies of single component vesicles (\citep{Vlahovska2007, misbah2006vacillating}) – the difference here is that we are examining a situation where $\boldsymbol{f}^{mem}$ takes a more complicated form (see section \ref{sec:dyn_equations}).

To solve Stokes flow, we first define the fundamental basis sets for Lamb’s solution – i.e., a series solution using vector spherical harmonics.  We will use the notation in \citet{vlahovska2005deformation, Blawzdziewicz2000}.  First, let us define the three vector spherical harmonics as follows
    \begin{equation} \label{eq:vec_harmonics}
        \boldsymbol{y}_{lm0} = \left[l(l+1)\right]^{-1/2}r\nabla Y_{lm} \qquad \boldsymbol{y}_{lm1} = i\boldsymbol{\hat{r}}\times \boldsymbol{y}_{lm0} \qquad \boldsymbol{y}_{lm2} = \boldsymbol{\hat{r}}Y_{lm}
\end{equation}

Using these quantities, one can define the velocity basis sets $\boldsymbol{u^{\pm}}_{lmq}$ ($q = 0, 1, 2$) where $\pm$ represents growing or decaying harmonics, respectively: 

\begin{subequations}
    \begin{equation}
        \boldsymbol{u^{-}}_{lm0} = \frac{1}{2}r^{-l}(2-l+lr^{-2})\boldsymbol{y}_{lm0} + \frac{1}{2}r^{-l}[l(l+1)]^{1/2}(1-r^{-2})\boldsymbol{y}_{lm2}
    \end{equation}

    \begin{equation}
        \boldsymbol{u^{-}}_{lm1} = r^{-l-1}\boldsymbol{y}_{lm1}
    \end{equation}

    \begin{equation}
        \boldsymbol{u^{-}}_{lm2} = \frac{1}{2}r^{-l}(2-l)\left(\frac{l}{l+1}\right)^{1/2}(1-r^{-2})\boldsymbol{y}_{lm0} + \frac{1}{2}r^{-l}(l+(2-l)r^{-2})\boldsymbol{y}_{lm2}
    \end{equation}
\end{subequations}

\begin{subequations}
    \begin{equation}
        \boldsymbol{u^{+}}_{lm0} = \frac{1}{2}r^{l-1}(-(l+1)+(l+3)r^{2})\boldsymbol{y}_{lm0} - \frac{1}{2}r^{l-1}[l(l+1)]^{1/2}(1-r^{2})\boldsymbol{y}_{lm2}
    \end{equation}

    \begin{equation}
        \boldsymbol{u^{+}}_{lm1} = r^{l}\boldsymbol{y}_{lm1}
    \end{equation}

    \begin{equation}
        \boldsymbol{u^{+}}_{lm2} = \frac{1}{2}r^{l-1}(3+l)\left(\frac{l+1}{l}\right)^{1/2}(1-r^{2})\boldsymbol{y}_{lm0} + \frac{1}{2}r^{l-1}((l+3)-(l+1)r^{2})\boldsymbol{y}_{lm2}
    \end{equation}
\end{subequations}

In terms of these basis sets, the velocity fields inside and outside the vesicle are:

\begin{equation} \label{eq:vel_field}
    \boldsymbol{u}^{out} = \sum_{lmq}c^{\infty}_{lmq} \boldsymbol{u}^{+}_{lmq}(\boldsymbol{x}) + \sum_{lmq}( c_{lmq}^{-} - c^{\infty}_{lmq} ) \boldsymbol{u}^{-}_{lmq}(\boldsymbol{x})
\end{equation}


\begin{equation}
    \boldsymbol{u}^{in} = \sum_{lmq}c_{lmq}^{+}\boldsymbol{u}^{+}_{lmq}(\boldsymbol{x})
\end{equation}

In the above equation, $c_{lmq}^{\infty}$ are the coefficients associated with the far field $\boldsymbol{u}^{\infty} = y \hat{\boldsymbol{x}}$, which correspond to $c_{2 \pm2 0}^{\infty} = \mp i \sqrt{\pi/5}$, $c_{2 \pm2 2}^{\infty} = \mp i \sqrt{2\pi/15}$ $c_{101}^{\infty} = i \sqrt{2\pi/3}$.  The coefficients $(c_{lmq}^{-} - c_{lmq}^{\infty})$ and $c_{lmq}^{+}$ are associated with the disturbance fields that decay outside the vesicle and grow inside the vesicle.  

Solving the Stokes equations thus reduces to solving the unknown coefficients associated with the disturbance velocity fields, as well as the surface tension on the interface.  For each mode $(l, m)$, there are seven unknowns to solve:  
\begin{equation*}
    (c_{lm0}^{+}, c_{lm1}^{+}, c_{lm2}^{+}, c_{lm0}^{-}, c_{lm1}^{-}, c_{lm2}^{-}, \sigma_{lm})
\end{equation*}
These are found by applying continuity of velocity across the surface $\boldsymbol{u}^{out} = \boldsymbol{u}^{in}$ at $r = 1$ (3 equations), membrane incompressibility $\nabla_s \cdot \boldsymbol{u}^{in} = 0$ at $r = 1$ (1 equation), and traction balance $\boldsymbol{f}^{hydro} = \boldsymbol{f}^{mem}$ at $r = 1$ (3 equations), where $\boldsymbol{f}^{mem}$ is Taylor expanded to $O(\Delta^{1/2})$ on the unit sphere. 

Once one performs this procedure, one then applies the kinematic boundary condition (Eq \ref{eq:kinematic_bc_vector}).  Doing so yields the differential equation for the shape mode $f_{lm}$ of the vesicle:

\begin{equation} \label{eq:shape_eqn}
\frac{d f_{lm}}{dt} = \frac{im}{2} f_{lm} + c_{lm2}^{+}
\end{equation}

The first term on the right hand comes from the rigid body rotation from shear flow, while the next term comes from the extensional deformation, which depends on the shape modes $f_{lm}$, concentration modes $q_{lm}$, and isotropic surface tension $\sigma_0$.  Appendix \ref{appB} presents the algebraic details to solve for $c_{lm2}^{+}$.  The final results are presented here:

\begin{equation} \label{eq:c_lm2}
    c_{lm2}^{+} = \lambda^{-1}C_{lm} + \lambda^{-1}\chi^{-1} \left[ ( A_l + B_l \sigma_0) f_{lm} + D_l q_{lm} \right]
\end{equation}
where the coefficients ($A_l, B_l, C_{lm}, D_l$) are listed in Eqs. (\ref{eq:coefs_final_expression})-(\ref{eq:d}) in the Appendix.  These depend on the the mode number $(l,m)$ as well as the dimensionless quantities discussed in Table \ref{tbl:Dimensionless_parameter_range}.

We note that the above expression (Eq \ref{eq:c_lm2}) depends on the isotropic surface tension $\sigma_0$, which one obtains by applying the the constant area constraint given by:
\begin{equation}\label{eq:area_constraint}
    A = 4\pi + \Delta,
\end{equation}
where the expression for excess area $\Delta$ is given in Eq. \eqref{eq:area_constraint_1}.  One analytically solves for $\sigma_0$ by taking the time derivative of Eq. \eqref{eq:area_constraint} and substituting the differential equation Eq. \eqref{eq:shape_eqn} for $\frac{d f_{lm}}{dt}$.  


\subsection{Solving Cahn-Hilliard equation} \label{sec:solve_CH}

To determine how the concentration modes $q_{lm}$ evolve over time, we solve the Cahn-Hilliard equation.  In the perturbative limit $\Delta \ll 1$, we will Taylor expand all quantities to $O(\Delta^{1/2})$ on the unit sphere except the terms coming from the double-well potential (i.e., $\frac{1}{2}aq^2 + \frac{1}{4}bq^4$), where we keep all higher order terms in the chemical potential.  The reason why we perform this task is that such higher order terms are needed to have a quartic free energy expression, which is necessary to have two-phase coexistence with a tie line. We note that similar approximations have been applied for equilibrium studies of multicomponent vesicles \citep{Luo_Maibaum_2020}.  In the cited study, the free energies are Taylor expanded to quadratic order in the shape and concentration modes except the double well term, where modes are kept to quartic order.  This yields tractions and chemical potentials to the same order of accuracy as the current study.

The Cahn-Hilliard equation becomes:

\begin{equation}\label{eq:CH_reduced}
\begin{split}
\frac{dq_{lm}}{dt} = \frac{im}{2}q_{lm} - \frac{1}{Pe} &[ \Lambda_{lm} l(l+1) + Cn^{2}l^{2}(l+1)^{2}q_{lm}   \\ 
&+2\alpha Cn^{2}\beta (l-1)l(l+1)(l+2)f_{lm} ] 
\end{split}
\end{equation}
where $\Lambda_{lm} = \int_{\Omega}\left(aq+bq^{3}\right)Y_{lm}d\Omega$ is the chemical potential from the double-well potential, projected onto spherical harmonics by integrating over a unit sphere.  This term is weakly nonlinear as discussed above.

\subsection{Numerical procedure} \label{sec:numerics}
The final equations we solve are equations (\ref{eq:shape_eqn})-(\ref{eq:area_constraint}) for the vesicle shape, and Eq (\ref{eq:CH_reduced}) for the concentration.  These are two coupled, nonlinear ordinary differential equations.  

When solving these equations, we decompose the shape and concentration modes into real and imaginary parts
\begin{equation}
    f_{lm} = f'_{lm}+if''_{lm} \qquad q_{lm} = q'_{lm}+iq''_{lm}
\end{equation} 
and perform an ODE solve for the components ($f'_{lm}, f''_{lm}, q'_{lm}, q''_{lm}$). We only solve for components with $m \geq 0$ since for the radius to be real, $f'_{l-m} = (-1)^m f'_{lm}$ and $f''_{l-m} = (-1)^{m+1} f''_{lm}$, with similar relations holding for the concentration.  Since we need to ensure that the constraint of area conservation is satisfied $A = 4\pi + \Delta$, utilizing regular solvers is not feasible. We have developed a numerical method to solve the ordinary differential equations as mentioned in  appendix \ref{sec:AppNumerical_Scheme}. After solving for the spectral coefficients, we reconstruct the vesicle shape and concentration fields using the spherical harmonic series in Eqs. \ref{eq:radius} and \ref{eq:order_param}.

\begin{figure}
  \centering{\includegraphics[width=0.7\textwidth]{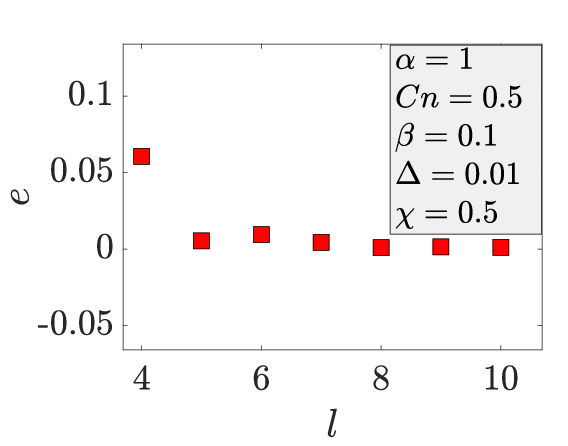}}
  \caption{Error convergence plots for $Cn=0.5$. The parameters for the simulations are $\alpha=1,\beta=0.1,\Delta=0.01,\chi=0.5$ The $y-$axis represents the ratio of the difference in the order parameter over the vesicle surface divided by the average magnitude of the order parameter of the case with most modes $l_{max} = 11$ }
\label{fig:Error_Convergence}
\end{figure}

{	For most of the simulations we perform, we choose a cutoff mode number between $l_{max} = 6$ and $l_{max} = 10$ when computing the order parameter.  We find that beyond this number, the dominant modes $0 \leq l \leq 4$ for shape and order parameter change less than 2 percent, and thus the trends discussed in the results section do not change appreciably.  Figure \ref{fig:Error_Convergence} shows an example of a convergence test.  Here, we computed the order parameter $q(\theta,\phi, t)$ using different cutoff modes, and compared the results against a simulation using a high cutoff mode.  The relative error we computed was $e = \int_S (\bar{q}_{approx} – \bar{q}_{true}) dA/\int_S \bar{q}_{true}dA$, where $\bar{q}_{true}$ is the time-averaged value at $(\theta,\phi)$ using the highest cutoff mode and $\bar{q}_{approx}$ is the time-averaged value at $(\theta, \phi)$ using the lower cutoff mode.  We are confident in our analysis for $Cn \sim 0.4-1$ which represents a physically realizable range of line tension values (see table \ref{tbl:Dimensionless_parameter_range}). For much smaller values of $Cn$ ($Cn \ll 1$), we expect that a spectral method may lead to inaccuracies as one will encounter the sharp interface limit that is hard to resolve owing to the Gibbs' phenomenon. }


\section{Single component vesicles}\label{sec:single_component_ves}
 When we have a single phospholipid making up the lipid bilayer, the bending stiffness is homogeneous over the entire surface. This gives us three primary descriptors to describe the deformation dynamics exhibited by the vesicle $(a)$ capillary number $\chi = \eta R^3 \dot{\gamma}/k_0$ given by the ratio of bending to flow forces, $(b)$ viscosity ratio $\lambda$, and (c) excess area $\Delta$.

When one inspects a single component vesicle under shear flow where the inner and outer viscosities are matched ($\lambda=2$), for a capillary number that is $\chi \sim O(\Delta^{1/2})$, the vesicle experiences tank-treading motion that is described by a stationary inclination angle close to $\pi/4$ \citep{misbah2006vacillating,Vlahovska2007,Deschamps2009,kantsler2006transition}. The explicit form of this inclination angle has been listed as
\begin{equation}
    \phi_{0} = \frac{\pi}{4} - \frac{(9+23\lambda\Delta^{1/2})}{16\sqrt{30\pi}}
\end{equation}

As one increases the viscosity ratio beyond a critical value $\lambda_{c} \sim O(\frac{1}{\Delta^{1/2}})$, the tank-treading behaviour turns into a vacillated breathing motion and eventually into tumbling. The main difference between vacillated breathing and tumbling is the out-of-plane deformation (i.e., $f_{20}$ mode), as described by Vlahovska and others \citep{Vlahovska2007} as well as Danker and co-workers \citep{misbah2006vacillating}. A very high viscosity ratios, rigid body tumbling dominates thereby causing the $f_{20}$ mode to be nearly zero.





\begin{figure}
  \centering
  {\includegraphics[width=0.6\textwidth]{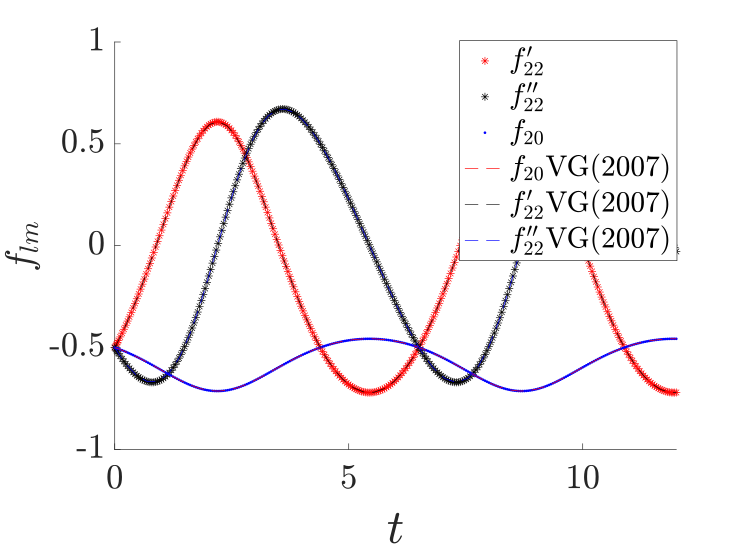}}
  \caption{Comparison of vesicle shape profiles against single component results (VG(2007)-\cite{Vlahovska2007}). The parameters are $Cn=0,\tilde{a}=0,\tilde{b}=0,\alpha=0,\beta=0,\lambda=20,\chi=0.6,\Delta=0.2$. {The initial condition is $f'_{22}(0)=-\sqrt{\frac{2\Delta\pi}{15}},f''_{22}(0)=-\sqrt{\frac{2\Delta\pi}{15}},f_{20}(0)=-\sqrt{\frac{2\Delta\pi}{15}}$} }
\label{fig:Comparison_SingleComponent}
\end{figure}

We validate our semi-analytical formulation (Eqs. \eqref{eq:shape_eqn} and \eqref{eq:CH_reduced}) by comparing against published, single component vesicle results.  In our simulations, we set $Cn=0,\alpha=0,\beta=0, \tilde{a}=0,\tilde{b}=0$ so that tractions arising from multicomponent lipids are zero. We present these results in figure \ref{fig:Comparison_SingleComponent} for $\lambda=20$. In these simulations, the initial condition is $f'_{22}(0)=-\sqrt{\frac{2\Delta\pi}{15}},f''_{22}(0)=-\sqrt{\frac{2\Delta\pi}{15}},f_{20}(0)=-\sqrt{\frac{2\Delta\pi}{15}}$. We plot $f_{lm}$ versus time and see that there is an exact agreement between the results of this study and previous studies \citep{Vlahovska2007}.

One can find phase diagrams corresponding to the tank-treading (TT), vacillated breathing (VB), and tumbling (TB) modes in previous publications \citep{misbah2006vacillating}. In the subsequent sections, we inspect the motion of multicomponent vesicles under shear flow to delineate the phase and deformation dynamics, while providing theoretical and experimental benchmarks wherever possible.


\section{Multicomponent vesicles}\label{sec:multicomponent_results}
The primary difference between single and multicomponent vesicles arises from the coupling between composition and the shape dynamics. This coupling arises from $(a)$ an inhomogeneous bending rigidity $\kappa_{c}$ over the vesicle, and $(b)$ advection-coarsening dynamics of different phases on the vesicle surface. 


We pick our primary descriptor to be the Peclet number $Pe = R^{2}\dot{\gamma}/\nu$ which describes the ratio of characteristic coarsening to flow times. This parameter has a very large range, depending on the relative physical parameters discussed in table \ref{tbl:Dimensionless_parameter_range}. In this paper, we inspect two ranges of values: $(a)$ large $Pe$ ($Pe \gg 1$) $(b)$ intermediate $Pe$ ($Pe \sim O(1)$). In the following sections, unless specified, the average order parameter $q_{0} = 0$, and the viscosities of the inner and outer fluid will be matched ($\lambda=2$). 

\subsection{$Pe \gg 1$ results}\label{sec:results_HighPe}
\subsubsection{Theory, $Pe \rightarrow \infty$}

When $Pe \to \infty$, the diffusive coarsening effects are negligible and the phospholipids are convected by the background shear flow.  Setting $Pe^{-1} = 0$ in Eq \ref{eq:CH_reduced} yields:
\begin{equation}
\frac{d q_{lm}}{dt} = \frac{im}{2} q_{lm}
\end{equation}
and hence the solution is oscillatory:  $q_{lm}(t) = q_{lm}(0) \exp(imt/2)$.  The shape of the vesicle is then given by Eqs (\ref{eq:shape_eqn})-(\ref{eq:area_constraint}) using the above expression for $q_{lm}$.

A previous study by \citep{gera_salac_spagnolie_2022} examined the shape of a 2D, multicomponent vesicle with no coarsening dynamics.  They found that the ($l=2$, $m = \pm 2$) subspace gives the dominant behaviour for the system, since in this situation the concentration field interacts with the imposed shear flow.  We will follow a similar setup and see what occurs for a 3D system, which to our knowledge has not been done before.  Let us examine the equatorial plane of the vesicle ($\theta = \pi/2$) and decompose the shape and concentration as follows (using the notation of \citep{gera_salac_spagnolie_2022}):

\begin{subequations} \label{eq:r_and_q_Pe_infty}
\begin{equation}
r(\theta = \pi/2, \phi, t) = 1 + \epsilon a_2(t) \cos(2\phi) + \epsilon b_2(t) \sin(2\phi)
\end{equation}
\begin{equation}
q(\theta = \pi/2, \phi, t) = \epsilon  \cos(2\phi + t)
\end{equation}
\end{subequations}
where $a_2$ and $b_2$ are time dependent coefficients to be determined, and $\epsilon = \Delta^{1/2}$ is the small parameter in this problem.  In terms of the coefficients ($f_{lm}, q_{lm}$) discussed in the paper, $f_{2\pm2} = 2 \epsilon \sqrt{\frac{2\pi}{15}} \left( a_2 \mp i b_2 \right)$ and $q_{2\pm2} = 2 \epsilon \sqrt{\frac{2\pi}{15}} \exp(\pm it)$.  If we define a modified capillary number as $C = \chi/\epsilon$ and modified time $\tau = t/\epsilon$, we find the leading-order shape dynamics in Eq (\ref{eq:shape_eqn}) for $C \sim O(1)$ and $\tau \sim O(1)$ to be:


 \begin{equation} \label{eq:ODE_Pe_infty_eta_1}
 \begin{split}
\frac{d a_2}{d \tau} &= -\Theta(a_2, b_2, , \epsilon \tau, C, \lambda, \beta) b_2 \\
\frac{d b_2}{d \tau} &= \Theta(a_2, b_2, , \epsilon \tau, C, \lambda, \beta) a_2
\end{split}
\end{equation}
These sets of ODEs take the same form as the 2D case, except that the expression $\Theta$ is different due to the geometry of the problem.  Apppendix \ref{sec:algebra_Pe_infty} shows the algebra to obtain $\Theta$ -- here we state the final results.  For the 3D system, we obtain:

\begin{equation} \label{eq:eta_3D}
    \Theta_{3D} = \frac{512 \pi}{5(9 + 23\lambda)} \left[ \frac{5}{8} a_2 + \beta C^{-1} \left( a_2 \sin(\epsilon \tau) + b_2 \cos(\epsilon \tau) \right) \right] 
\end{equation}
while for the 2D system, we obtain:
\begin{equation} \label{eq:eta_2D}
    \Theta_{2D} = \frac{3 \pi}{2\lambda} \left[ a_2 + \beta C^{-1} \left( a_2 \sin(\epsilon \tau) + b_2 \cos(\epsilon \tau) \right) \right] 
\end{equation}
where in the 2D case the small parameter $\epsilon$ is related to the excess length as $\epsilon = \Delta_{2D}^{1/2}$ (see Appendix \ref{sec:algebra_Pe_infty}).  The expressions for $\Theta$ depend on the viscosity ratio $\lambda$, reduced area $\epsilon^2 = \Delta$, and a lumped bending stiffness parameter $\beta C^{-1} = \beta \Delta^{1/2}/\chi$.  Results are discussed below.

\subsubsection{Dynamics and phase diagrams for theory}

The system of equations (\ref{eq:ODE_Pe_infty_eta_1}) admit a periodic solution for the shape modes $a_2(t)$ and $b_2(t)$.  Two different behaviours are seen:  \\

\begin{figure}
  \centerline{\includegraphics[width=0.9\textwidth]{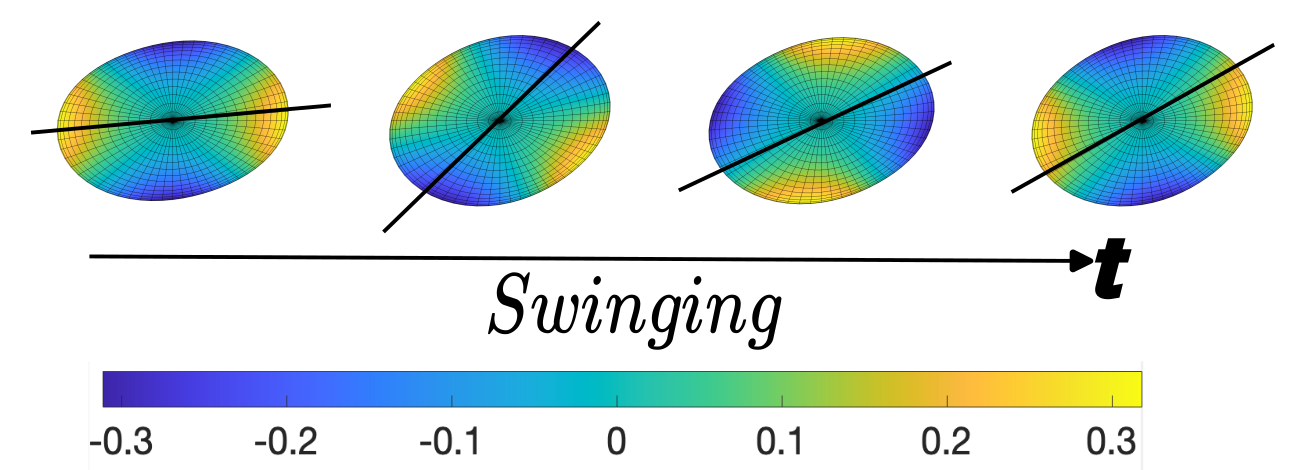}}
  \caption{ Swinging visualization. The parameters are $Pe=10000,a=-1,b=1,Cn=0.5,\alpha=1,\beta=0.5,\Delta=0.1,\chi=0.5$. The blue phase represents the softer phase and the yellow phase represents the stiffer phase. See supplementary video for visualization. }
\label{fig:Swinging_Viz_HighPe}
\end{figure}

\begin{enumerate}
    \item \underline{Swinging} -- The inclination angle undergoes weak oscillations around a fixed angle $\phi$, while the concentration field rotates along the membrane (for visualization, see figure \ref{fig:Swinging_Viz_HighPe}).  
    \item \underline{Tumbling} -- The vesicle shape and concentration field undergo rigid body rotation  (for visualization and detailed discussion, see figure \ref{fig:Tumbling_Visualization_V1}).\\
\end{enumerate}

\begin{figure}
  \centerline{\includegraphics[width=0.9\textwidth]{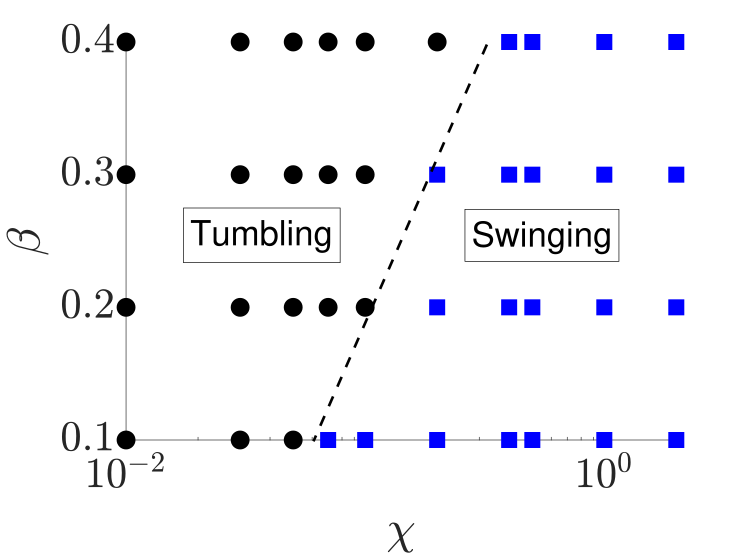}}
  \caption{ Phase diagram for $Pe \to \infty$ limit at $\Delta=0.0001$ and $\lambda = 2$. }
\label{fig:HighPe_PhaseDiagram}
\end{figure}

For a fixed viscosity ratio $\lambda$ and excess area $\Delta$, the transition between the two behaviours is determined by the dimensionless quantity $\beta C^{-1} = \beta \Delta^{1/2}/\chi$.  Below a critical value of $\beta C^{-1}$, swinging is seen to occur, while tumbling is seen for values above this critical value.  Figure \ref{fig:HighPe_PhaseDiagram} shows a plot where we vary $\beta$,$\chi$ to understand the dynamics of the vesicle with $\beta C^{-1}$ . We have plotted this phase diagram for $\Delta =0.0001$, since this theory is valid for nearly spherical vesicles. We see that the critical conditions exhibit similar trends for 3D and 2D vesicles, although their quantitative values might be different \citep{gera_salac_spagnolie_2022}. The explanation given by the authors can be summarized as follows: (a) At low background shear rates, the vesicle elongates in the direction of the softer phase in a quasi-steady manner. These regions have a higher curvature. In order to match the phase to the curvature, the vesicle undergoes rigid body tumbling. (b) On the other hand, when the shear rates are high, the viscous stresses dominate over the elastic stresses thereby pushing the stiffer phase past the high curvature region. In order to account for this phase-shape mismatch, the vesicle undergoes a swinging motion.  The variation with respect to the viscosity ratio  $\lambda$ has been discussed later in section \ref{sec:PhaseDiagram}.

\begin{figure}
\centering
\begin{subfigure}{0.49\textwidth}
    {\includegraphics[width=\textwidth]{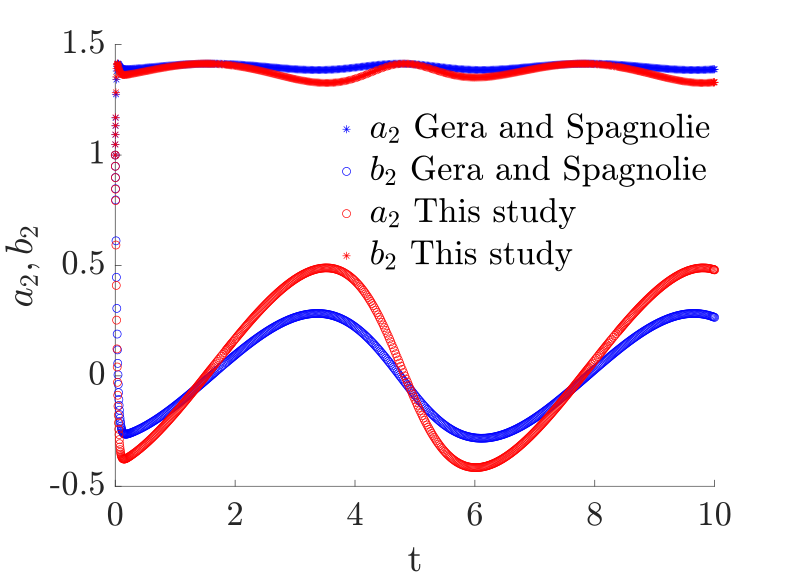}}
    \captionsetup{font=normalsize,labelfont={bf,sf}}
        \caption{Swinging $\beta=1,\chi/\Delta^{1/2}=5$, $\lambda = 2$}
        \label{fig:Swinging_Comparison}
\end{subfigure}
  \begin{subfigure}{0.49\textwidth}
    {\includegraphics[width=\textwidth]{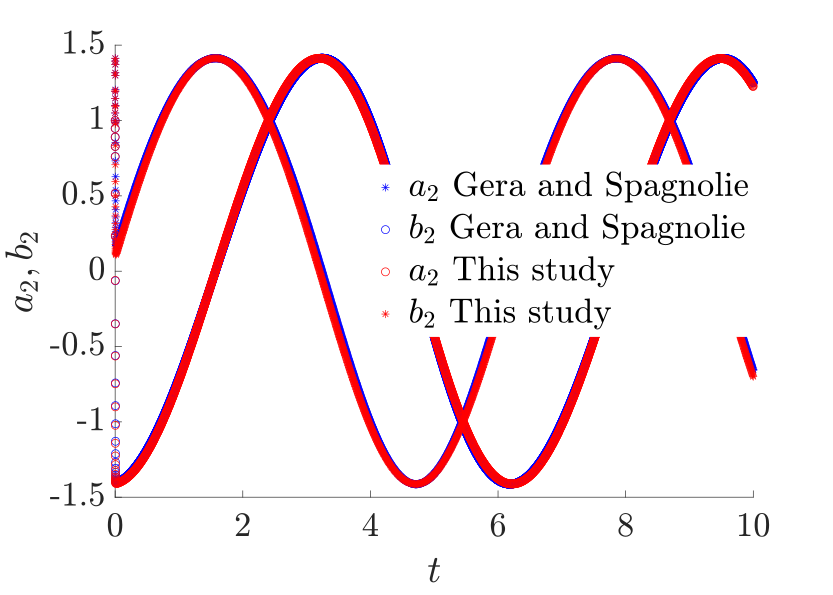}} \captionsetup{font=normalsize,labelfont={bf,sf}}
        \caption{Tumbling $\beta=1,\chi/\Delta^{1/2}=0.2$, $\lambda = 2$}
        \label{fig:Tumbling_Comparison}
\end{subfigure}
  \caption{$Pe \rightarrow \infty$ results comparison with \cite{gera_salac_spagnolie_2022} at $\Delta=0.0001$ by solving Eq. \eqref{eq:ODE_Pe_infty_eta_1}. }
\label{fig:High_Pe_flm_Comparison}
\end{figure}

Figure \ref{fig:High_Pe_flm_Comparison} shows a plot for $a_2(t)$ and $b_2(t)$ for swinging and tumbling vesicles, comparing the 3D and 2D theories in the limit $Pe \rightarrow \infty$.  As mentioned, the agreement is not exact due to multiple pre-factors entering the 3D analysis compared to the 2D case.  However, the agreement for tumbling is nearly exact in both cases. The discrepancy primarily arises due to the differences in expressions for $\Theta_{2D}$ and $\Theta_{3D}$.

\subsubsection{Comparison with numerical simulations}
\begin{figure}
\centering
\begin{subfigure}{0.49\textwidth}
\includegraphics[width=\textwidth]{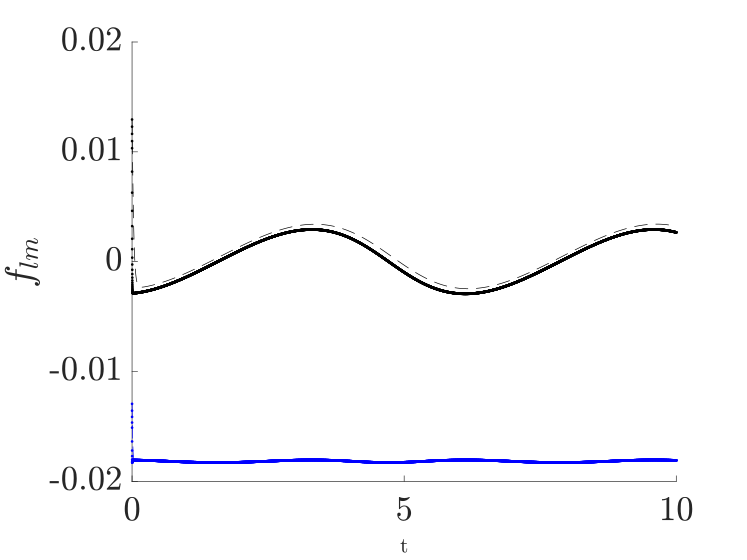}
\captionsetup{font=normalsize,labelfont={bf,sf}}
    \caption{$f_{lm}$ vs $t$}
    \label{fig:High_Pe_flm_Swinging}
\end{subfigure}
\begin{subfigure}{0.49\textwidth}
\includegraphics[width=\textwidth]{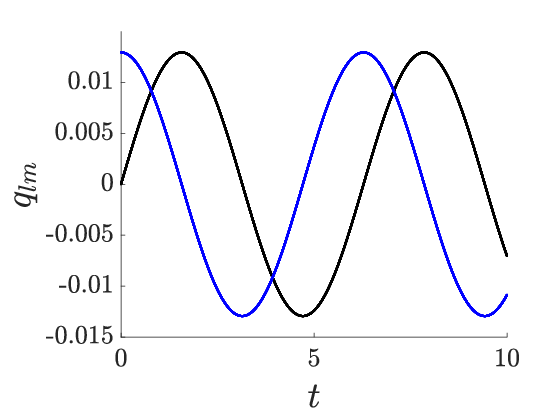}
\captionsetup{font=normalsize,labelfont={bf,sf}}
    \caption{$q_{lm}$ vs $t$}
    \label{fig:High_Pe_phi_Swinging}
\end{subfigure}

\caption{High $Pe$ swinging motion for $\alpha=0.1,Cn=0.1,\beta=0.5,Pe=1000,\chi=0.05,q_{0}=0,a=-1,b=1,\Delta=0.0001$. The dashed lines represent the full numerical solution of Eqs. \eqref{eq:shape_eqn},\eqref{eq:CH_reduced} whereas the dots represent $f'_{22},f''_{22},q_{22}',q''_{22}$ for the $Pe \rightarrow \infty$ theory Eqs. \eqref{eq:r_and_q_Pe_infty}, \eqref{eq:ODE_Pe_infty_eta_1}.}
\label{fig:High_Pe_Swinging}
\end{figure}

We perform full numerical simulations to check the validity of $Pe \rightarrow \infty$ theory discussed previously.  Figure \ref{fig:High_Pe_Swinging} shows a case where swinging occurs at $Pe=1000$ ($\lambda = 2, \alpha=0,\chi =0.2,\Delta=0.01,Cn=0,\beta=0.5$).  The solid lines indicate the theory (Eq. \ref{eq:ODE_Pe_infty_eta_1}), while the symbols are the full numerical simulations (solving Eqs.  (\ref{eq:shape_eqn})-(\ref{eq:area_constraint}) and (\ref{eq:CH_reduced})).  The shape modes $f_{lm}$ are plotted as well as the order parameter $q'_{22},q''_{22}$.  In the simulation, we start with an initial condition that has both deformations in and out of plane -- $f'_{22}(0)=2\sqrt{\frac{2\pi\Delta}{15}}\sqrt(\frac{15}{64\pi}),f''_{22}(0)=-2\sqrt{\frac{2\pi\Delta}{15}}\sqrt(\frac{15}{64\pi}),f_{20}(0)=0$. We observe that the out-of-plane $f_{20}$ deformations rapidly decay to zero and the simulations closely match the theory. We have generally seen this behaviour hold for $Pe \sim O(100)$ and above. Another interesting observation that is important is the validity of the $Pe \to \infty$ expressions (Eq. \eqref{eq:ODE_Pe_infty_eta_1}). These expressions only hold true when $\Delta \ll 1$. As our $\Delta$ increases, the rotational term in Eq. \eqref{eq:shape_eqn} becomes significant and starts influencing the dynamics as well. Hence, the approximation given by expressions Eq. \eqref{eq:ODE_Pe_infty_eta_1} are not accurate as $\Delta$ increases.

\begin{figure}
\centering
\begin{subfigure}{0.49\textwidth}
\includegraphics[width=\textwidth]{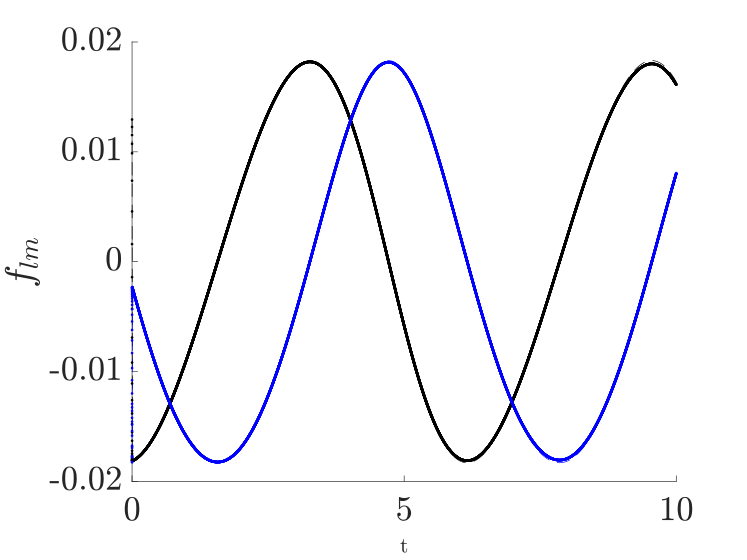}
\captionsetup{font=normalsize,labelfont={bf,sf}}
    \caption{$f_{lm}$ vs $t$}
    \label{fig:High_Pe_flm_Tumbling}
\end{subfigure}
\begin{subfigure}{0.49\textwidth}
\includegraphics[width=\textwidth]{figs/q_comparison_Tumbling.png}
\captionsetup{font=normalsize,labelfont={bf,sf}}
    \caption{$q_{lm}$ vs $t$}
    \label{fig:High_Pe_phi_Tumbling}
\end{subfigure}

\caption{High $Pe$ tumbling motion for $\alpha=0.1,Cn=0.1,\beta=0.5,Pe=1000,\chi=0.001,q_{0}=0,a=-1,b=1,\Delta=0.0001$. The dashed lines represent the full numerical solution of Eqs. \eqref{eq:shape_eqn},\eqref{eq:CH_reduced} whereas the dots represent $f'_{22},f''_{22},q_{22}',q''_{22}$ for the $Pe \rightarrow \infty$ theory Eqs. \eqref{eq:r_and_q_Pe_infty}, \eqref{eq:ODE_Pe_infty_eta_1}. }
\label{fig:High_Pe_Tumbling}
\end{figure}

Similarly, figure \ref{fig:High_Pe_Tumbling} shows a simulation where tumbling occurs at $Pe = 1000$ ($\lambda = 2, \chi = 0.01, \beta = 0.5$).  This is characterized by $f'_{22}(t)\sim \cos(t),f''_{22}\sim \sin(t)$. The initial condition is $f'_{22}(0)=\sqrt{\frac{2\pi\Delta}{15}},f''_{22}(0)=-\sqrt{\frac{2\pi\Delta}{15}},f_{20}(0)=\sqrt{\frac{2\pi\Delta}{15}}$. We can observe this motion in figure \ref{fig:High_Pe_Tumbling} where we plot $f_{lm}$ and $\phi_{i}$ vs $t$.

\subsection{Intermediate $Pe$}
In this section, we focus on vesicles undergoing motion for a $Pe\sim O(1-10)$.  Under this regime, the flow  and coarsening timescales are comparable. 

\subsubsection{General observations}
In the previous section ($Pe \gg 1$), we found that the vesicle motion depends on the viscosity ratio ($\lambda$), excess area $(\epsilon^2 = \Delta$), and the lumped bending stiffness parameter $\beta C^{-1} = \beta \chi/\epsilon$.  When $Pe \sim O(1)$, the coarsening dynamics also become important, and thus additional parameters to consider are:\\

\begin{enumerate}
\item \underline{Line tension parameter $\alpha  = k_0/\gamma^2$}, which indicates the relative bending to line tension energy.
\item \underline{Cahn number $Cn = \epsilon^{width}/(\sqrt{2}R)$}, which is related to the interface width compared to the vesicle size.  This is also related to line tension.\\
\end{enumerate}

The coarsening dynamics give rise to a wider range of vesicle shape behaviours compared to the previous section.  Below summarizes the behaviours observed in the long time limit when one excites the $l=2$ concentration mode.\\

\begin{figure}
  \centerline{\includegraphics[width=\textwidth]{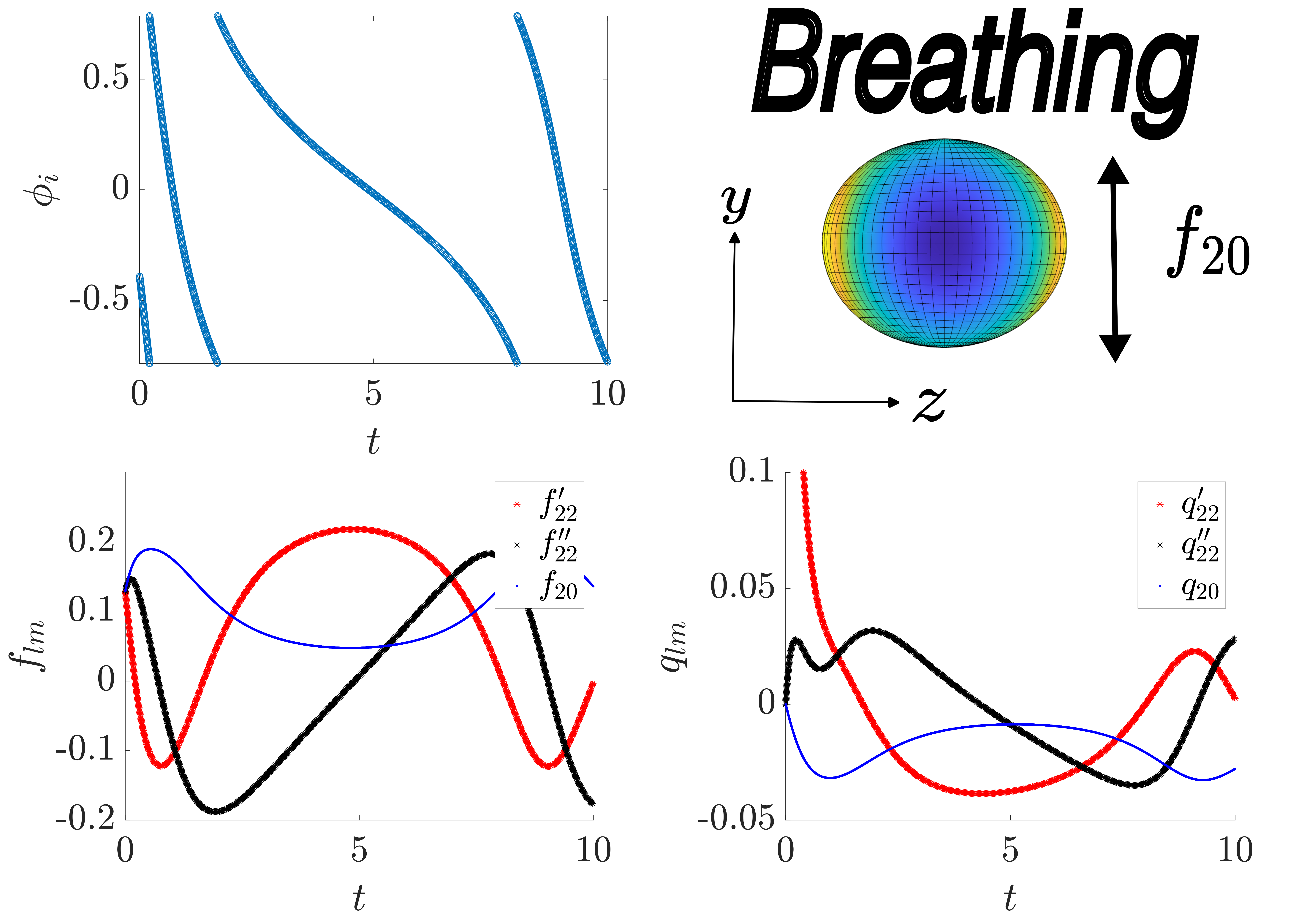}}
  \caption{ Breathing. The parameters are $\chi = 0.01,\alpha=1,Cn=0.8,\beta = 0.1,\lambda=10,Pe=5,\Delta=0.1$. The blue phase represents the softer phase and the yellow phase represents the stiffer phase. }
\label{fig:Breathing_Viz}
\end{figure}

\begin{enumerate}
\item \underline{Tank treading}.  Here, both the vesicle’s shape and concentration field are stationary.  The vesicle is ellipsodal and fixed at a steady inclination angle $\phi$, and the concentration field is frozen (figure \ref{fig:Tank_qlm_delta001}). 
\begin{figure}
  \centerline{\includegraphics[width=\textwidth]{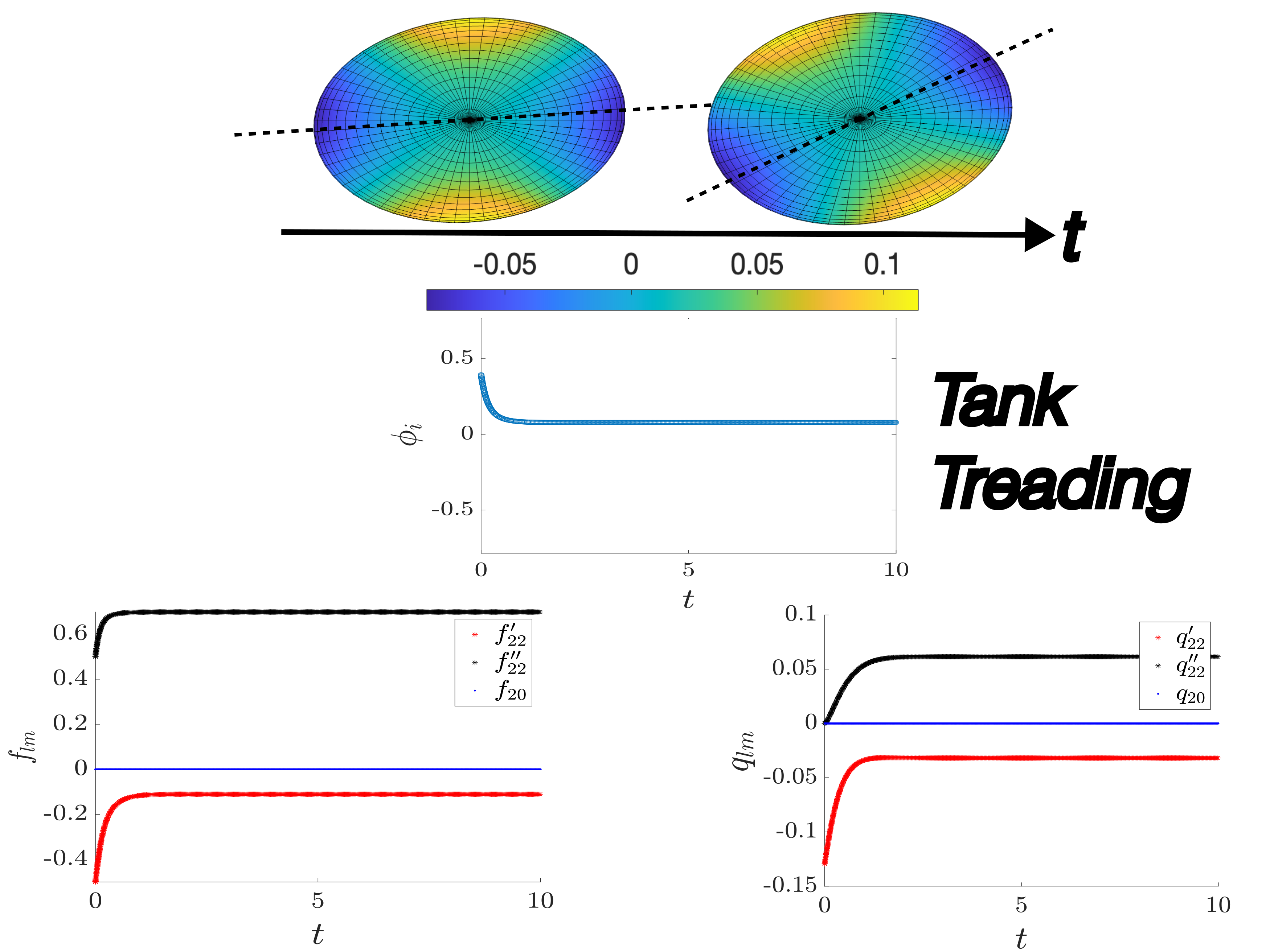}}
  \caption{Tank treading. The parameters are $\chi = 0.1,\alpha=1,Cn=1,\beta = 0.1,\lambda=2,Pe=1,\Delta=0.1$. The blue phase represents the softer phase and the yellow phase represents the stiffer phase. See supplementary video for visualization.}
\label{fig:Tank_qlm_delta001}
\end{figure}
\item \underline{Tumbling}.  The vesicle’s shape and concentration field exhibit rigid body rotation (figure \ref{fig:Tumbling_Visualization_V1}).

\begin{figure}
  \centerline{\includegraphics[width=1.1\textwidth]{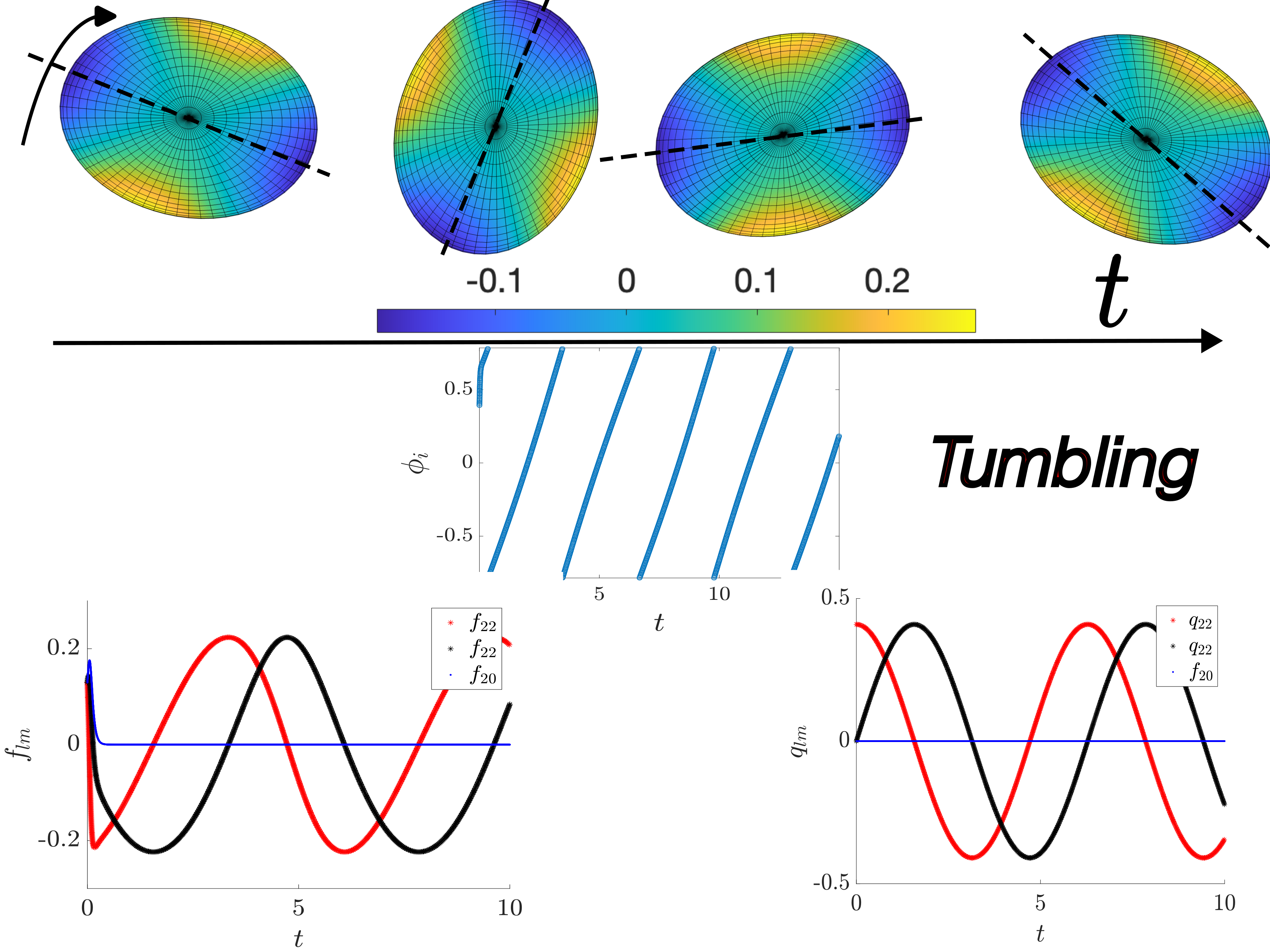}}
  \caption{ Tumbling for parameters $\alpha=1,\beta=0.1,Cn=0.4,\chi=0.01,Pe=5,\lambda=2,\Delta=0.1$. The blue phase represents the softer phase and the yellow phase represents the stiffer phase. See supplementary video for visualization}
\label{fig:Tumbling_Visualization_V1}
\end{figure}

\item \underline{Phase treading}.  The vesicle’s inclination angle is steady at a fixed angle $\phi$, and the concentration rotates around the membrane.  This behaviour looks similar to swinging, except the orientation does not oscillate considerably in magnitude (figure \ref{fig:PhaseTreading_Visualization_V1}).

\begin{figure}
  \centerline{\includegraphics[width=1.1\textwidth]{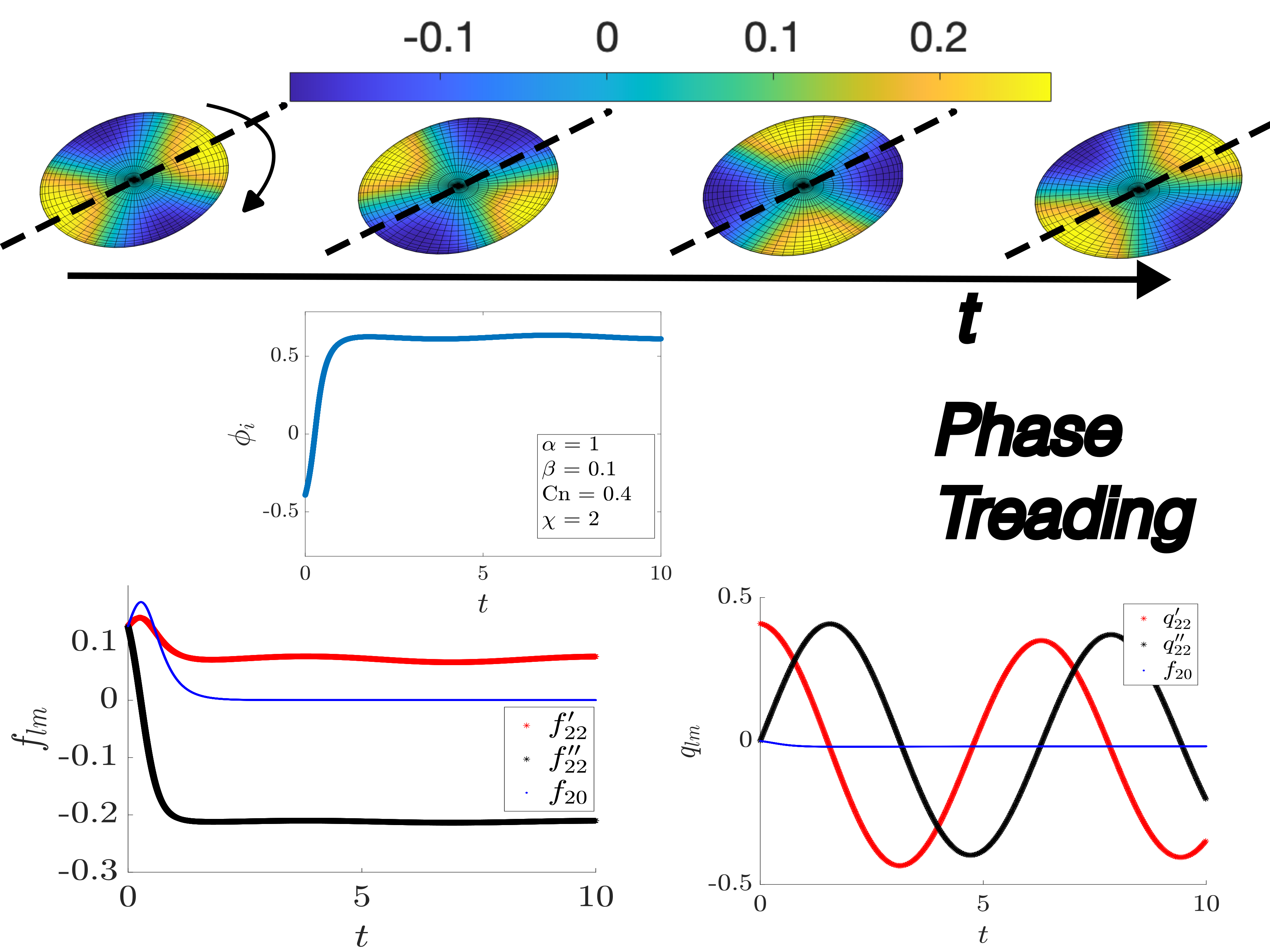}}
  \caption{Phase treading for parameters $\alpha=1,\beta=0.1,Cn=0.4,\chi=2,Pe=5,\Delta = 0.1,\lambda=2$. The blue phase represents the softer phase and the yellow phase represents the stiffer phase. See supplementary video for visualization. }
\label{fig:PhaseTreading_Visualization_V1}
\end{figure}

\item \underline{Vacillated breathing/trembling}.  The vesicle performs oscillatory rotations about an axis in the flow-gradient plane while there are significant deformations in the flow-vorticity plane represented by the $f_{20}$ mode. (figure \ref{fig:Breathing_Viz}).\\
\end{enumerate}

Figures \ref{fig:Tank_qlm_delta001}-\ref{fig:Breathing_Viz} show snapshots of these behaviours, as well as typical plots for the shape modes $f_{lm}$, concentration modes $q_{lm}$, and inclination angle $\phi$.  Movies are also provided in the Supporting Information.

\begin{figure}
\centering
\begin{subfigure}{0.49\textwidth}
\includegraphics[width=\textwidth]{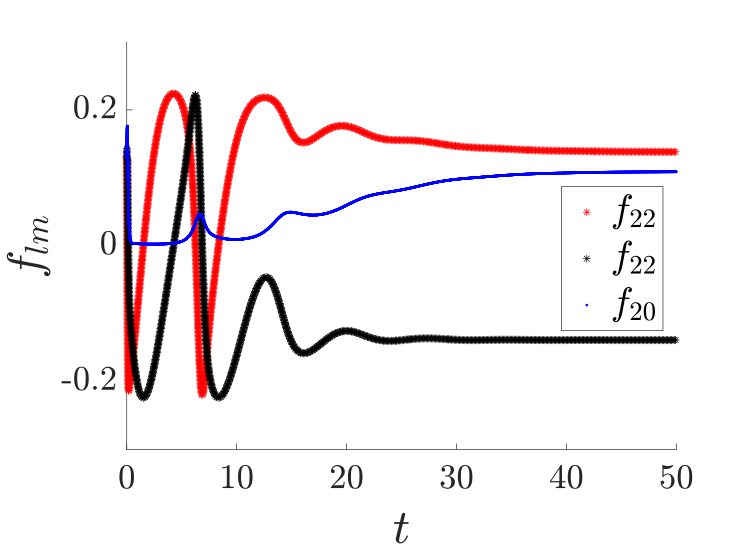}
\captionsetup{font=normalsize,labelfont={bf,sf}}
    \caption{$f_{lm}$ vs $t$}
    \label{fig:Oscillatory_Dampening_flm}
\end{subfigure}
\begin{subfigure}{0.49\textwidth}
\includegraphics[width=\textwidth]{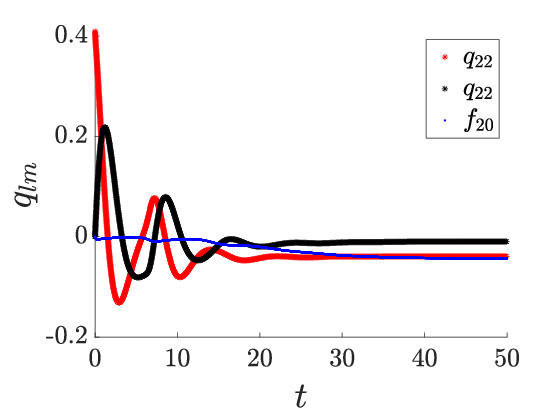}
\captionsetup{font=normalsize,labelfont={bf,sf}}
    \caption{$q_{lm}$ vs $t$}
    \label{fig:Oscillatory_Dampening_qlm}
\end{subfigure}

\caption{An example of tumbling motion dampening to give rise to tank-treading. The parameters are $\lambda=2,\beta=0.1,Cn=0.5,Pe=5,\Delta=0.1,\chi=0.01$}
\label{fig:Oscillatory_Dampening}
\end{figure}

Occasionally, we have observed situations where the vesicle undergoes tumbling motion before the motion dampens out at very long times to tank treading behaviour. This has been noted in figure \ref{fig:Oscillatory_Dampening}. We believe this is not a numerical instability but arises from the nonlinear mode-mixing term $bq^{3}$ in the Cahn Hilliard equation. It is important therefore to understand what is the timescale at which these motions are observed experimentally. In figure \ref{fig:Oscillatory_Dampening}, we see that the tumbling motion persists up to $t\sim 10$ which translates to a physical time of $10{\dot{\gamma}}^{-1}$ that could range from $O(1)-O(10)$ $s$. Shear flow experiments are usually limited by issues like photo-bleaching or escape of the vesicle from the field-of-view. Hence, we believe in order to characterize the motion, the timescale of observation is very important and must be reported for a phase diagram.  For the other dynamical regimes listed above, we do not observe a change of behaviour at very long times $(t \sim O(100))$.



\subsubsection{Phase diagrams}\label{sec:PhaseDiagram}

This section provides phase diagrams to quantify general trends for vesicle dynamics.  Since there is a large region of phase space to explore, we will isolate the effect of a few variables mentioned in section \ref{sec:dim_quantities}. 

\begin{figure}
\centering
\begin{subfigure}{0.49\textwidth}
\includegraphics[width=\textwidth]{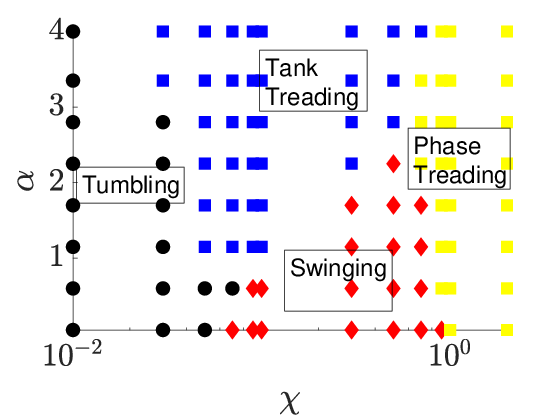}
\captionsetup{font=normalsize,labelfont={bf,sf}}
    \caption{$Cn=0.4$}
    \label{fig:Phase_Plot_Cn04_alpha_chi}
\end{subfigure}
\begin{subfigure}{0.49\textwidth}
\includegraphics[width=\textwidth]{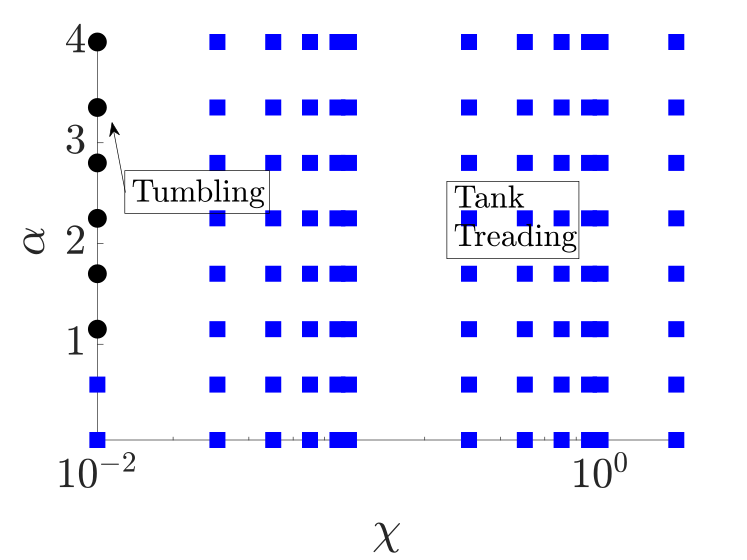}
\captionsetup{font=normalsize,labelfont={bf,sf}}
    \caption{$Cn=0.48$}
    \label{fig:Phase_Plot_Cn048_alpha_chi}
\end{subfigure}

\caption{Phase diagram for $\chi$ vs $\alpha$ at $\Delta=0.1,Pe=5,\beta=0.1,\lambda=2$. }
\label{fig:Phase_Plot_alpha_chi}
\end{figure}

The first set of variables we will understand are the capillary number ($\chi$) and line tension parameter ($\alpha$).  We will choose a moderately deflated vesicle ($\Delta = 0.1$) with matched interior and exterior fluid viscosity ($\lambda = 2$) and zero average order parameter ($q_0 = 0$), and choose $O(1)$ values for the other parameters:  $Pe = 5, \beta = 0.1$.  Figure \ref{fig:Phase_Plot_alpha_chi} shows the types of motion observed by the vesicle for different values of capillary number $\chi$ and line tension parameter $\alpha$, when we start the initial condition with only the $l = 2$ shape and concentration modes excited. 
 {Two different interface widths are examined:  $Cn = 0.4$ (more sharp) and $Cn = 0.48$ (less sharp).} The initial condition in all these simulations is $f'_{22}(0) = 2\sqrt{\frac{\Delta}{32}},f''_{22}(0) = 2\sqrt{\frac{\Delta}{32}},f_{20}(0) = 2\sqrt{\frac{\Delta}{32}},q'_{22}(0) = 2\sqrt{\frac{\Delta}{32}}, q_{22}''(0)=q_{20}(0) = 0$.  The regimes are reported over time window $t \approx 50$.

We will now explain the reasons behind these phase diagrams: In the first case where we see the relatively sharper interface $Cn=0.4$ (figure \ref{fig:Phase_Plot_Cn04_alpha_chi}), we see that at small $\chi$ values, the vesicle shows tumbling motion. This value of $\chi$ lowers the value of variable $C = \chi/\Delta^{1/2}$ in the shape equations \eqref{eq:shape_eqn}-\eqref{eq:c_lm2} since $f_{lm}\sim O(\Delta^{1/2})$.  This causes the vesicle to elongate in the softer phase direction, thereby driving a phase-curvature mismatch, leading to tumbling motion, very similar to what is seen in the high $Pe$ limit (see section \ref{sec:results_HighPe}). 

When we increase this $\chi$ value, depending on how stiff the vesicle is, given by $\alpha$, we can get tank-treading or swinging motion. When $\alpha$ is small, the $\frac{im}{2}q_{lm}$ term in the Cahn Hilliard equation \eqref{eq:CH_reduced} drives the oscillatory behaviour of $q_{lm}$ since the curvature-phase coupling term $\alpha\beta Cn^{2} f_{lm}$ is small and cannot dampen the oscillations easily. This translates to the shape equation \eqref{eq:shape_eqn}-\eqref{eq:c_lm2} showing oscillatory motion due to the $q_{lm}$ term, but the $\chi^{-1}$ prefactor in \eqref{eq:c_lm2} dampens these oscillations leading to swinging motion.  When the $\alpha$ value increases ($\alpha>1$), we see that the curvature-phase coupling term $\alpha\beta Cn^{2}f_{lm}$ in the Cahn-Hilliard equation dampens the oscillatory behaviour caused by the background shear flow, leading to tank-treading. 

At very large $\chi$ values, the $\chi^{-1}$ prefactor $\chi^{-1}$ in Eq. \eqref{eq:c_lm2} dampens the shape oscillations, leading to a tank-treading like motion, but the phases are convected by the background shear. This causes phase treading motion, which has been explored by authors previously \citep{gera2018three}.

As the vesicle inteface becomes more diffuse ($Cn$ increases), we see that the vesicle shows behaviour more similar to single-component vesicles. This is seen in  figure \ref{fig:Phase_Plot_Cn048_alpha_chi} where the $Cn=0.48$. At small $\chi$ values, the term $\chi/\Delta^{1/2}$ is small thereby driving tumbling motion as stated before. As $\chi$ increases, the $f_{lm}$ oscillations are dampened out in the shape equations \eqref{eq:shape_eqn}-\eqref{eq:c_lm2} and the $q_{lm}$ oscillations in the Cahn Hilliard equation \eqref{eq:CH_reduced} are dampened due to the line tension penalty term $Cn^{2}l^{2}(l+1)^{2}q_{lm}$ coming into play. As this $Cn$  increases further, the vesicle will largely show tank-treading behaviour, which is expected for single component vesicles with a matched viscosity ($\lambda=2)$.
 

\begin{figure}
  \centerline{\includegraphics[width=0.8\textwidth]{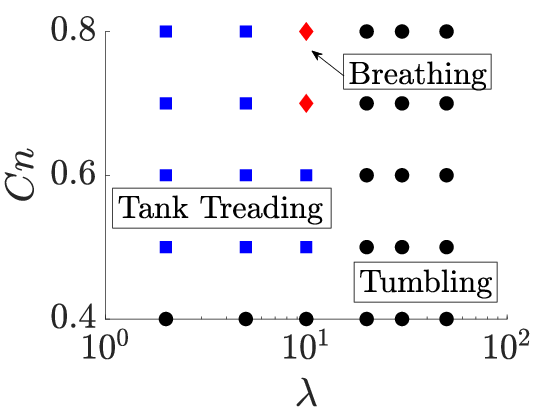}}
  \caption{Phase plot for $Cn$ vs viscosity ratio $\lambda$. The parameters for simulations are $\beta=0.1,\alpha=1,a=-1,b=1,Pe=5,\Delta=0.1,\chi=0.01$}
\label{fig:Cn_Lambda_Plot}
\end{figure}

Figure \ref{fig:Cn_Lambda_Plot} discusses the effect of viscosity ratio $\lambda$ on the phase behaviour.  We find the effect of $\lambda$ depends greatly on the sharpness of the interface, parameterized by the Cahn number $Cn$. When $Cn$ is on the higher end (i.e., diffuse interface), single component behaviour prevails, leading to the tank-treading $\rightarrow$ breathing $\rightarrow$ tumbling transitions as discussed in previous publications \citep{misbah2006vacillating,Vlahovska2007}. As $Cn$ reduces, we see some interesting trends.  For sharp interfaces ($Cn \leq 0.4$), we only see tumbling behaviour, primarily due to the fact that the simulations in the phase diagram were performed at small capillary numbers ($\chi/\Delta^{1/2} \ll 1$), where this behaviour is only observed as discussed previously (see figure \ref{fig:Phase_Plot_Cn04_alpha_chi}). As we go to $Cn=0.5$, there is a bifurcation in $\lambda$ where there is a tank-treading to tumbling transition. At some point $Cn \sim 0.6-0.7$, there is an additional bifurcation that leads to breathing motion as seen in figure \ref{fig:Breathing_Viz}. A detailed explanation behind this motion is seen in \citep{misbah2006vacillating}.  The authors explain this motion by talking about the compromise between tank-treading and tumbling motion due to the transfer of torque to the tank-treading motion and elongation to the tumbling motion causing the vesicle to vacillate between tank-treading and tumbling.

In the following subsection, we discuss experimental considerations that could act as an aid in future studies. 
\subsubsection{Phase diagrams using experimental parameters}

In an experiment, one creates a vesicle dispersion and places it in an external shear flow.  The lipid composition is fixed, yielding constant material parameters for the membrane (e.g., bending stiffness, phase separation, and line tension parameters).  However, the dispersion is polydisperse in size $R$ and excess area $\Delta$.  Thus, in a given experiment, one adjusts the shear rate $\dot{\gamma}$ and visualizes the vesicle dynamics for many different particle sizes $R$ and excess areas $\Delta$.  In terms of the dimensionless numbers listed in Table \ref{tbl:Dimensionless_parameter_range}, the following parameters are fixed:
\begin{equation}
\lambda, \alpha, \beta, \tilde{a}, \tilde{b}, q_0
\end{equation}
while the following parameters can vary in an experiment:
\begin{equation}
\Delta, \chi, Cn, Pe
\end{equation}

Since one examines different values of three quantities $\dot{\gamma}$, $R$, and $\Delta$, the above dimensionless numbers are not independent of each other.  In particular, $\chi Cn/Pe = \frac{\eta \nu \gamma}{k_0^2}$ depends only on material properties and thus is a constant.  Thus, the appropriate phase space to consider is:
\begin{equation}
\Delta, \chi, Cn, Pe \qquad \text{subject to} \qquad \frac{\chi}{Pe} Cn = \text{const}
\end{equation}

\begin{figure}
\centering
\begin{subfigure}{0.48\textwidth}
\includegraphics[width=\textwidth]{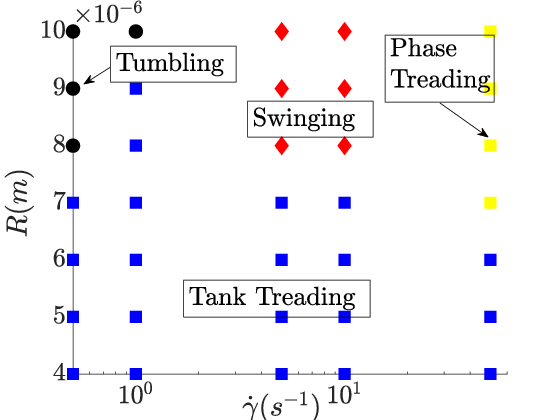}
\captionsetup{font=normalsize,labelfont={bf,sf}}
    \caption{$\Delta=0.1$}
    \label{fig:Experimental_Del01}
\end{subfigure}
\begin{subfigure}{0.51\textwidth}
\includegraphics[width=\textwidth]{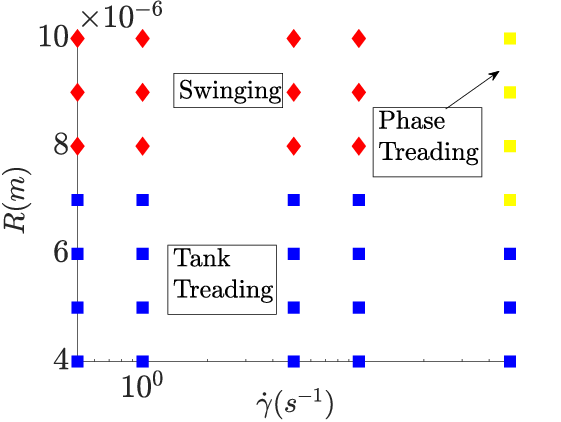}
\captionsetup{font=normalsize,labelfont={bf,sf}}
    \caption{$\Delta=0.01$}
    \label{fig:Experimental_Del001}
\end{subfigure}
\caption{Phase diagram for $R$ vs $\dot{\gamma}$ at $\beta=0.1,\lambda=2$. }
\label{fig:Experimental_Plot}
\end{figure}

Here, we will consider a typical experimental conditions listed in Table \ref{tbl:Physical_parameter_range} ($\eta=10^{-3}Pa\cdot s,k_{0}=10^{-18}J,k_{1}=10^{-19}J,\zeta_{0}=10^{-6}J,\gamma=3\times 10^{-9} J^{1/2}, \nu = 10^{-11} m^{2}/s $), and run shear flow simulations for shear rates $0.1 \leq \dot{\gamma} \leq 10$ $s^{-1}$, radius $5 < R < 25$ $\mu m$, and excess area $\Delta = 0.1,0.01$.  Figure \ref{fig:Experimental_Plot} shows phase diagrams for vesicle dyamics in the radius/shear rate space for the two values of the excess area $\Delta$. In the first figure \ref{fig:Experimental_Del01}, when the vesicle size is small, the Peclet number ($Pe$) and capillary number ($\chi$) are small leading to tank treading motion dominating. When $R<5 \mu m$, the value of $Cn$ is large $>0.6$ thereby causing the interface to be diffused and the vesicle to behave like an effective single-component vesicle due to the increased dampening in the shape and Cahn-Hilliard equations. This causes the vesicle to take up a stationary phase tank-treading motion. For larger vesicles $R\sim O(10)-O(100)$ $\mu$m, one observes different dynamical regimes. For $\Delta=0.1$ (figure \ref{fig:Experimental_Del01}), when the shear rate is low and $R$ is high, the capillary number to excess area ratio is small ($\chi/\Delta^{1/2} \ll 1$). As discussed in the previous section (section \ref{sec:PhaseDiagram}), this drives tumbling motion. When the shear rate $\dot{\gamma}$ increases, the capillary number increases, giving rise to swinging, followed phase treading motion. 
Similar observations can be made for the $\Delta=0.01$ case in figure \ref{fig:Experimental_Del001}. The only difference is seen in the low shear rate, large vesicle size limit. In this limit, the $\chi/\Delta^{1/2}$ ratio is larger thereby driving swinging motion instead of tumbling as seen before. In both the large vesicle cases, due to the $Pe\sim R^{2}$ scaling, we transition from the intermediate $Pe \sim O(1)$ to large $Pe$ ($Pe \gg 1$) case quite drastically.


We need to keep certain aspects in mind here. While performing experiments, the timescale of measurement is critical. We have observed, in certain cases, for intermediate $Pe$ values ($Pe \sim O(1)-O(10)$), vesicles whose long-time behaviour is tank treading exhibit tumbling at intermediate times before the oscillations are dampened out. If these oscillations dampen out within the measurement timescale, the motion would be measured as tank treading. If not, it would be tumbling. 


\section{Conclusion}\label{sec:discussion}

In this study, we explored the effect of shear flow around a multicomponent vesicle containing a ternary mixture of phospholipids and cholesterol molecules within the bilayer. We analyzed the small-deformation, weakly segregated limit in these vesicles. At a particular viscosity ratio, these vesicles show multiple deformation patterns, as opposed to their single-component counterparts. The multicomponent vesicles can exhibit tumbling, breathing, tank-treading, swinging, or phase treading, depending on the membrane stiffness inhomogeneity between phases, line tension between the phases, and coarsening time scales. When the $Pe \gg 1$, the phases are advected by the shear flow over the surface. The strength of the coupling with the shape deformations, given by $\beta \Delta^{1/2}/\chi$, determines the nature of this motion -- swinging or tumbling. The out-of-plane deformations do not matter significantly in this limit. The results show good qualitative agreement with previous two-dimensional studies on such vesicles at the $Pe \rightarrow \infty$ limit \citep{gera_salac_spagnolie_2022}.

When one inspects the intermediate $Pe$ limit ($Pe \sim O(1)-O(10)$), when the viscosity is matched, we see that when the interface is relatively sharper $Cn <0.45$, we see a variety of motions like tumbling, swinging, and phase-treading, and tank-treading. In the diffused interface limit, $Cn>0.5$, we see more of tank-treading behaviour as can be expected from a single component vesicle with matched viscosity. The vesicle behaves like an effective single component vesicle with stationary phases. We then provided a general overview of the dynamics by showing phase plots delineating the different dynamics that the vesicle experiences. We also discussed the effect of the viscosity ratio $\lambda$ that show transitions from tank-treading to breathing to tumbling. We concluded the study with an experimental phase diagram to aide an experimental study with operating parameters like $R$ and $\dot{\gamma}$ for a fixed set of lipids. 

We have been able to show a plethora of dynamics that the vesicle exhibits within the space space explored. While the weak-segregation, small-deformation limit eases up the mathematical rigor largely due to the exclusion of non-linear stresses, we believe it provides a great starting point for further semi-analytical treatments of these systems using higher-order theories.

\section*{Acknowledgments}
The authors would like to acknowledge support from National Science Foundation (grant 2147559-CBET).  

\section*{Declaration of Interests}
The authors report no conflicts of interest.

\appendix
\section{Spherical harmonics}\label{app:Spherical_Harmonics}

We define spherical harmonics $Y_{lm}$ using the following convention \citep{Jones1985}
\begin{equation}
    Y_{lm}(\theta, \phi) = \sqrt{\frac{(2l+1)(l-m)!}{4\pi(l+m)!}}(-1)^{m}P_{l}^{m}(\cos\theta)e^{im\phi},
\end{equation}
where $(\theta,\phi)$ are the polar and azimuthal angles, and $P_l^m$ are associated Legendre polynomials. The inner products of these spherical harmonics are orthonormal -- i.e., $\int Y_{lm} Y_{l'm'}^* d \Omega = \delta_{ll'} \delta_{mm'}$, where the integral is over the unit sphere and $*$ represents complex conjugate.  Furthermore, the following relationship holds: $Y_{l -m} = (-1)^m Y_{lm}^*$.

The vector spherical harmonics $\boldsymbol{y}_{lmq}$ in Eq (\ref{eq:vec_harmonics}) also have the following properties:
\begin{itemize}
    \item The different types of vector spherical harmonics are orthogonal by dot product:
        \begin{equation}
            \boldsymbol{y}_{lmq} \cdot \boldsymbol{y}_{lmq'} = 0 \quad q \neq q' ; \qquad \boldsymbol{y}_{lm2} \cdot \boldsymbol{y}_{l'm'q'} = 0 \quad q = \{0, 1\}
        \end{equation}
    \item They are orthogonal with respect to the inner product:
    \begin{subequations}
        \begin{equation}
                \int \boldsymbol{y}_{lmq} \cdot \boldsymbol{y}_{l'm'q'}^* d\Omega = \delta_{ll'} \delta_{mm'} \delta_{qq'} 
        \end{equation}
    \end{subequations}
        
\end{itemize} 




\section{Algebraic details for Stokes flow solution}\label{appB}

\subsection{Velocity boundary conditions}
Using the velocity fields in Eq (\ref{eq:vel_field}), we apply continuity of velocity ($\boldsymbol{u}^{in} = \boldsymbol{u}^{out}$) and membrane incompressibility ($\nabla_s \cdot \boldsymbol{u}^{in} = 0$) on the unit sphere $r = 1$.  This yields the following relationships:

\begin{equation}
    c_{lmq}^{+} = c_{lmq}^{-}, \quad q = {0,1,2}
\end{equation}

\begin{equation} \label{eq:mem_incompress}
c_{lm2}^{+} = \frac{1}{2} \sqrt{ l (l+1) } c_{lm0}^{+}
\end{equation}

\subsection{Hydrodynamic tractions on surface}

From the velocity fields in Eq (\ref{eq:vel_field}), one can compute the hydrodynamic stresses on the unit sphere $r = 1$.   These are written as follows:

\begin{equation}
    \boldsymbol{n} \cdot \boldsymbol{T}^{in} = (\lambda - 1) \sum_{lmq} \tau_{lmq}^{in} \boldsymbol{y}_{lmq} \qquad  \boldsymbol{n} \cdot \boldsymbol{T}^{out} = \sum_{lmq} \tau_{lmq}^{out} \boldsymbol{y}_{lmq}
\end{equation}

Using the notation and results from \citet{Blawzdziewicz2000}, the coefficients $\tau_{lmq}$ are related to the velocity coefficients $(c_{lmq}^{\infty}, c_{lmq}^{+}, c_{lmq}^{-})$ as follows: 

\begin{equation}\label{Appeq:tau_out}
    \tau^{out}_{lmq} = \sum_{q'=0}^{2}c^{\infty}_{lmq'}\left(\Theta^{+}_{q'q}-\Theta^{-}_{q'q}\right) + \sum_{q'=0}^{2}c_{lmq'}^{-}\Theta^{-}_{q'q}
\end{equation}

\begin{equation}\label{Appeq:tau_in}
    \tau^{in}_{lmq} =\sum_{q'=0}^{2}c_{lmq'}^{+}\Theta^{+}_{q'q},
\end{equation}
where 
\begin{equation}
    \Theta^{+}_{qq'}=\begin{bmatrix}
        (2l+1) & 0 & -3\left(\frac{l+1}{l}\right)^{1/2}\\ 
        0 & l-1 & 0 \\
        -3\left(\frac{l+1}{l}\right)^{1/2} & 0 & \left(2l+1+\frac{3}{l}\right)\\
    \end{bmatrix}
\end{equation}

\begin{equation}
    \Theta^{-}_{qq'}=\begin{bmatrix}
        (-2l-1) & 0 & 3\left(\frac{l}{l+1}\right)^{1/2}\\ 
        0 & -l-2 & 0 \\
        3\left(\frac{l}{l+1}\right)^{1/2} & 0 & \left(-2l-1-\frac{3}{l+1}\right)\\
    \end{bmatrix}
\end{equation}

\subsection{Membrane tractions on surface}

Using vector spherical harmonics (Eq \ref{eq:vec_harmonics}), we Taylor expand the membrane tractions to $O(\Delta^{1/2})$. 
 This yields:
 
 \begin{equation}
     \boldsymbol{f}^{mem} = {t}^{mem}_{lm0}\boldsymbol{y}_{lm0}+{t}^{mem}_{lm2}\boldsymbol{y}_{lm2}
 \end{equation}
where

\begin{equation}\label{eq:tmemlm0}
    {t}^{mem}_{lm0} = \left(\left[-2\beta - \frac{\mu_{0}}{\alpha Cn^{2}}\right]q_{lm} - \sigma_{lm}\right)\sqrt{l(l+1)}
\end{equation}
and

\begin{multline}\label{eq:tmemlm2}
    {t}^{mem}_{lm2} = 2\beta l(l+1) q_{lm} + \\
    \left(\frac{(l+2)(l-1)g_{0}}{\alpha Cn^{2}} + (1+\beta q_{0})(l-1)(l)(l+1)(l+2)+ \sigma_{0}(l-1)(l+2)\right)f_{lm} + 2\sigma_{lm}
\end{multline}

In the above equation, $g_{0} = \frac{1}{2}aq_{0}^{2} + \frac{1}{4}bq_{0}^{4}$ is the double well potential evaluated at the base state $q_0$, and $\mu_{0} = aq_{0} + bq_{0}^{3}$ is the chemical potential associated with this double-well potential.
The dimensionless quantities $\alpha, Cn, \beta$ are defined in table \ref{tbl:Dimensionless_parameter_range}, which correspond to the line tension parameter, Cahn number, and dimensionless bending stiffness difference.


\subsection{Solving equations}
In non-dimensional form, the traction boundary condition is $\chi \boldsymbol{f}^{hydro} = \boldsymbol{f}^{mem}$ at $r = 1$, where $\chi$ is the capillary number in table \ref{tbl:Dimensionless_parameter_range} and the tractions are given by Eqs  \ref{Appeq:tau_out}-\ref{eq:tmemlm2}.  This yields:

\begin{equation}\label{eq:membrane_balance_lm0}
    \chi\left[\tau^{out}_{lmq} - (\lambda-1)\tau^{in}_{lmq}\right] = t^{mem}_{lmq}
\end{equation}

The solenoidal component ($q = 1$) does not have a membrane traction ($t_{lm1}^{mem} = 0$), which yields the following condition:

\begin{equation}
c_{lm1}^{+} = c_{lm1}^{-} = c_{lm1}^{\infty}
\end{equation}
i.e., the rotational component of the flow moves simply as rigid body motion. The other two conditions on the traction balance yield the following relations:\\

\begin{itemize}
\item Tangential stress balance ($q = 0$):

\begin{equation}
    \chi c^{\infty}_{lm0}\frac{2(2l+1)}{\sqrt{l(l+1)}} - \chi c^{\infty}_{lm2}\frac{3(2l+1)}{l(l+1)} - \chi c_{lm0}^{+}\frac{3+\lambda(l-1)}{2\sqrt{l(l+1)}}= \left(2\beta + \frac{\mu_{0}}{\alpha Cn^{2}}\right)q_{lm} - \sigma_{lm}  
\end{equation}

\item Normal stress balance ($q = 2$):
\begin{equation}
\begin{split}
    &\frac{-3(2l+1)}{\sqrt{l(l+1)}} \chi c^{\infty}_{lm0}  + \frac{2l+1}{l(l+1)} (2l^2 + 2l + 3) \chi c^{\infty}_{lm2} \\
    &-\frac{1}{2\sqrt{l(l+1)}} \chi c_{lm0}^{+} \left( 3 + \lambda (l+1) (2l^2 + l - 3) \right) = t_{lm2}^{mem}  
\end{split}
\end{equation}

\end{itemize}

These equations give two equations for two unknowns ($\sigma_{lm}, c_{lm0}^{+}$).  Solving these equations and using the relation $c_{lm2}^{+} = \frac{1}{2}\sqrt{l(l+1)} c_{lm0}^{+}$ from membrane incompressibility (Eq \ref{eq:mem_incompress}) yield the expression in the main text: 

\begin{equation*}
    c_{lm2}^{+} = \lambda^{-1}C_{lm} + \lambda^{-1}\chi^{-1} \left[ ( A_l + B_l \sigma_0) f_{lm} + D_l q_{lm} \right]
\end{equation*}
where $C_{lm}$ is a term that depends on the external flow $\boldsymbol{u}^{\infty}$, while $A_l$, $B_l$, and $D_l$ are coefficients shown below:




\begin{subequations} \label{eq:coefs_final_expression}

    \begin{equation}
        C_{lm} = \frac{[c_{lm0}^{\infty}\sqrt{l(l+1)}(2l+1) + c^{\infty}_{lm2}(4l^{3}+6l^{2}-4l-3)]}{d(\lambda,l)}
    \end{equation}

    \begin{equation}
        A_{l} = -\frac{(l-1)l(l+1)(l+2)}{d(\lambda,l)}\left[ (1+\beta q_{0})l(l+1) + \frac{g_{0}}{\alpha Cn^{2}}\right]
    \end{equation}

    \begin{equation}
        B_{l} = -\frac{(l-1)l(l+1)(l+2)}{d(\lambda,l)}
    \end{equation}
    
    \begin{equation}
        D_l = -\frac{2(l-1)l(l+1)(l+2)\beta}{d(\lambda,l)}  - \frac{2l(l+1)\mu_{0}}{\alpha Cn^{2}d(\lambda,l)}
    \end{equation}
    
\end{subequations}
where 

\begin{equation} \label{eq:d}
    d(\lambda,l) = \frac{9}{\lambda} + (2l^{3}+3l^{2}-5)
\end{equation}

\section{Numerical scheme to solve ordinary differential equations} \label{sec:AppNumerical_Scheme}

We start with the following ordinary differential equations for the deformation and order parameter dynamics:

\begin{equation} 
\frac{d f_{lm}}{dt} = \frac{im}{2} f_{lm} +  \lambda^{-1}C_{lm} + \lambda^{-1}\chi^{-1} \left[ ( A_l + B_l \sigma_0) f_{lm} + D_l q_{lm} \right]
\end{equation}

\begin{equation}
\begin{split}
\frac{dq_{lm}}{dt} = \frac{im}{2}q_{lm} - \frac{1}{Pe} &[ \Lambda_{lm} l(l+1) + Cn^{2}l^{2}(l+1)^{2}q_{lm}   \\ 
&+2\alpha Cn^{2}\beta (l-1)l(l+1)(l+2)f_{lm} ] 
\end{split}
\end{equation}

In the above equations, $f_{lm},q_{lm}$, and $\sigma_{0}$ are unknown. We start by splitting $f_{lm},q_{lm}$ into their real and imaginary parts -- e.g., $f_{lm}=f'_{lm}+if''_{lm}$ -- while keeping in mind that $f_{l-m} = (-1)^{|m|}f^{*}_{l|m|}$. For the deformation, this gives us $(m\geq 0)$:

\begin{equation}\label{flm_app_real}
    \frac{df'_{lm}}{dt} = -\frac{m}{2}f''_{lm} + \lambda^{-1}\Re(C_{lm}) + \lambda^{-1}\chi^{-1}\left[\Re(\left(\left(A_{l}+B_{l}\sigma_{0}\right)f_{lm}\right))\right] + D_{l}q'_{lm}
\end{equation}

\begin{equation}\label{flm_app_imag}
    \frac{df''_{lm}}{dt} = \frac{m}{2}f'_{lm} + \lambda^{-1}\Im(C_{lm}) + \lambda^{-1}\chi^{-1}\left[\Im(\left(\left(A_{l}+B_{l}\sigma_{0}\right)f_{lm}\right))\right] + D_{l}q''_{lm}
\end{equation}
where $\Re(),\Im()$ represent the real and imaginary parts of variables respectively. Similarly, for the order parameter, we get the following $(m\geq 0)$:

\begin{equation}
\begin{split}
\frac{dq'_{lm}}{dt} = \frac{-m}{2}q''_{lm} - \frac{1}{Pe} & [\Re(\Lambda_{lm}) l(l+1) + Cn^{2}l^{2}(l+1)^{2}q'_{lm}   \\ 
&+2\alpha Cn^{2}\beta (l-1)l(l+1)(l+2)f'_{lm}] 
\end{split}
\end{equation}

\begin{equation}
\begin{split}
\frac{dq''_{lm}}{dt} = \frac{m}{2}q'_{lm} - \frac{1}{Pe} & [\Im(\Lambda_{lm}) l(l+1) + Cn^{2}l^{2}(l+1)^{2}q''_{lm}   \\ 
&+2\alpha Cn^{2}\beta (l-1)l(l+1)(l+2)f''_{lm}] 
\end{split}
\end{equation}

Now that we have the expressions for the real and imaginary components, we proceed in the following manner:

At time $t=t_{i}$, we perform a first-order Euler predictor-corrector scheme to march the vesicle shape \eqref{flm_app_real}-\eqref{flm_app_imag}.  We first take a predictor step while neglecting the $\sigma_{0}$ terms

\begin{equation}
    \frac{\tilde{f}'_{lm}-{f'}^{i}_{lm}}{\Delta t} = -\frac{m}{2}{f''}^{i}_{lm} + \lambda^{-1}\Re(C_{lm}) + \lambda^{-1}\chi^{-1}\left[\left(\left(A_{l}{f'}^{i}_{lm}\right)\right)\right] + D_{l}{q'}^{i}_{lm}
\end{equation}

\begin{equation}
    \frac{\tilde{f}''_{lm}-{f''}^{i}_{lm}}{\Delta t} = \frac{m}{2}{f'}^{i}_{lm} + \lambda^{-1}\Im(C_{lm}) + \lambda^{-1}\chi^{-1}\left[\left(\left(A_{l}{f''}^{i}_{lm}\right)\right)\right] + D_{l}{q''}^{i}_{lm}
\end{equation}
where $\tilde{f}'_{lm},\tilde{f}''_{lm}$ are intermediate predictor values. We now proceed to take another step, now considering only the $\sigma_{0}$ portion of Eqs \ref{flm_app_real},\ref{flm_app_imag}. We choose an appropriate guess value for $\sigma_{0}$ here.

\begin{equation}
    \frac{\tilde{\tilde{f}}'_{lm}-\tilde{f}'_{lm}}{\Delta t} = \lambda^{-1}\chi^{-1}\Re(B_{l}\sigma_{0}f_{lm})
\end{equation}

\begin{equation}
    \frac{\tilde{\tilde{f}}''_{lm}-\tilde{f}''_{lm}}{\Delta t} = \lambda^{-1}\chi^{-1}\Im(B_{l}\sigma_{0}f_{lm})
\end{equation}

The choice of $\sigma_{0}$ may not satisfy the constraint $A/R^{2} = 4\pi + \Delta$. In order to do so, we use the $fzero$ function in MATLAB to impose the constraint :

\begin{equation}
      \sum_{l,m\geq 0}{\frac{(l-1)(l+2)}{2} \left({\tilde{\tilde{f}}'_{lm}}^{2}+{\tilde{\tilde{f}}''_{lm}}^{2}\right)} - \Delta = 0
\end{equation}

This gives the optimum value of $\sigma_{0}$. The reason for doing this step is to ensure that the dynamical behaviour of the non-linear equations do not drift the differential-algebraic equation system away from the constraint\citep{ascher1994stabilization}. This method seems to perform better than simply imposing $\sum_{lm}(l-1)(l+2)\frac{df_{lm}}{dt}f^{*}_{lm} = 0$ to calculate $\sigma_{0}$ analytically. We set the tolerance of $fzero$ to be $10^{-8}$. Once we obtain $\sigma^{opt}_{0}$, we simply calculate the deformation at $t=t_{i+1}$.

\begin{equation}
    \frac{{f'_{lm}}^{i+1}-\tilde{f}'_{lm}}{\Delta t} = \lambda^{-1}\chi^{-1}\Re(B_{l}\sigma^{opt}_{0}f_{lm})
\end{equation}

\begin{equation}
    \frac{{f''_{lm}}^{i+1}-\tilde{f}''_{lm}}{\Delta t} = \lambda^{-1}\chi^{-1}\Im(B_{l}\sigma^{opt}_{0}f_{lm})
\end{equation}

We are left with the convective Cahn-Hilliard equation that uses a simpler procedure. We split $aq+bq^{3} = (a-2)q+bq^{3}+2q$ as shown in previous studies \citep{yoon2020fourier}. We treat the linear terms implicitly and non-linear terms implicitly while using the $f_{lm}$ data from $t=t_{i+1}$

\begin{equation}
\begin{split}
\frac{{q'_{lm}}^{i+1} - {q'_{lm}}^{i+1} }{\Delta t} = \frac{-m}{2}{q''_{lm}}^{i+1} - \frac{1}{Pe} & [\Re({\Lambda'_{lm}}^{i})l(l+1) + 2{q'_{lm}}^{i+1}l(l+1) + Cn^{2}l^{2}(l+1)^{2}{q'_{lm}}^{i+1}   \\ 
&+2\alpha Cn^{2}\beta (l-1)l(l+1)(l+2){f'_{lm}}^{i+1}] 
\end{split}
\end{equation}

\begin{equation}
\begin{split}
\frac{{q''_{lm}}^{i+1} - {q'_{lm}}^{i+1} }{\Delta t} = \frac{m}{2}{q'_{lm}}^{i+1} - \frac{1}{Pe} & [\Im({\Lambda'_{lm}}^{i})l(l+1) + 2{q''_{lm}}^{i+1}l(l+1) + Cn^{2}l^{2}(l+1)^{2}{q''_{lm}}^{i+1}   \\ 
&+2\alpha Cn^{2}\beta (l-1)l(l+1)(l+2){f''_{lm}}^{i+1}] 
\end{split}
\end{equation}

where $\Lambda'_{lm} = \int_{\Omega}\left((a-2)q+bq^{3}\right)Y_{lm}d\Omega$

This gives us the deformation and order parameter at the $t_{i+1}$ time step. This method helps us conserve the area at every time step. This becomes very important while dealing with $Pe\sim O(1)$ cases.

\section{Algebra, $Pe \rightarrow \infty$ theory} \label{sec:algebra_Pe_infty}
We examine Eqs (\ref{eq:shape_eqn}-\ref{eq:c_lm2}) and use the expressions $f_{22} = 2 \epsilon \sqrt{\frac{2\pi}{15}} \left( a_2 - i b_2 \right)$ and $q_{22} = 2 \epsilon \sqrt{\frac{2\pi}{15}} \exp(it)$.   If we define a modified capillary number as $C = \chi/\epsilon$ and modified time $\tau = t/\epsilon$, the system of equations at leading order for $C \sim O(1), \tau \sim O(1)$ are:
 
\begin{equation} \label{eq:ODE_Pe_infty}
\begin{split}
\frac{d a_2}{d \tau} &= \lambda^{-1} C^{-1} \left[ (A_2 + B_2 \sigma_0) a_2 + D_2 \cos(\epsilon \tau) \right]\\
\frac{d b_2}{d \tau} &= \lambda^{-1} C^{-1} \left[ (A_2 + B_2 \sigma_0) b_2 - D_2 \sin(\epsilon \tau) \right]  - \lambda^{-1} \frac{1}{2i} \sqrt{\frac{15}{2\pi}} C_{22} 
\end{split}
\end{equation}
 In the above equation, $A_2, B_2, C_{22}$, and $D_2$ are coefficients defined in Eqs. (\ref{eq:coefs_final_expression})-(\ref{eq:d}) for $l = 2, m = 2$.  The surface tension $\sigma_0$ is a Lagrange multiplier to enforce the surface area constraint -- i.e., $a_2^2 + b_2^2 = \frac{15}{32\pi}$.  We solve for $\sigma_0$ using the expression $a_2 \frac{d a_2}{d \tau} + b_2 \frac{d b_2}{d \tau} = 0$, and plug it back into the above ODE (\ref{eq:ODE_Pe_infty}) to yield: 

 \begin{equation} \label{eq:ODE_Pe_infty_eta}
 \begin{split}
\frac{d a_2}{d \tau} &= -\Theta(a_2, b_2, C, \lambda, \beta) b_2 \\
\frac{d b_2}{d \tau} &= \Theta(a_2, b_2, C, \lambda, \beta) a_2
\end{split}
\end{equation}
 where $\Theta$ is:

 \begin{equation}
 \Theta(a_2, b_2, C, \lambda, \beta) = -\frac{1}{a_2^2 + b_2^2} \left[ \frac{1}{2i} \sqrt{\frac{15}{2\pi}} \lambda^{-1} C_{22} a_2 + \lambda^{-1} C^{-1} D_2 \left( a_2 \sin(\epsilon \tau) + b_2 \cos(\epsilon \tau) \right)  \right]
 \end{equation}
Substituting the expressions for $C_{22}$ and $D_2$ (i.e.,  $C_{22} = -60 \lambda i \sqrt{\frac{2\pi}{15}} \left( 9 + 23 \lambda \right)^{-1}$,  $D_2 = -48 \lambda \beta \left( 9 + 23 \lambda \right)^{-1}$ for $q_0 = 0$), as well as the constraint $a_2^2 + b_2^2 = \frac{15}{32\pi}$ yields expression (\ref{eq:eta_3D}) in the main text.

In \citep{gera_salac_spagnolie_2022}, the authors obtained the same form of the differential equation as Eq (\ref{eq:ODE_Pe_infty_eta}) for $a_2$ and $b_2$ for a 2D vesicle, when the radius is written as $r = 1 + \epsilon a_2(t) \cos(2\phi) + \epsilon b_2 \sin(2\phi)$.  However, the expression for $\Theta$ is different due the fact the vesicle is 2D rather than 3D.  Their expression is (using the notation in our paper):
 \begin{equation}
 \Theta_{2D} = \frac{1}{\lambda} \frac{1}{Q^2} \left[  \left( 1 + \beta C^{-1} \sin(\epsilon \tau) \right) a_2 + \beta C^{-1} \cos(\epsilon \tau) b_2  \right]
 \end{equation}
where $Q$ is a constant equal to $Q^2 = \frac{1}{3}\sum_{n=1}^{\infty}(n^2 - 1) (a_n^2 + b_n^2)$.  If one defines a 2D excess length parameter as $\Delta_{2D} = L/R - 2\pi$, where $L$ is the membrane length and area $A = \pi R^2$, one obtains:

\begin{equation}
    \Delta_{2D} = \frac{3\pi}{2} \epsilon^2 Q^2
\end{equation}
Keeping consistent with the notation used in our paper, if we define the small parameter $\epsilon = \Delta_{2D}^{1/2}$, this yields $Q^2 = \frac{2}{3\pi}$, which gives the final expression written in the main text (Eq (\ref{eq:eta_2D})).

\bibliographystyle{jfm}

\end{document}